\documentclass[review,1p,times]{elsarticle} 

\usepackage{hyperref}
\usepackage{lineno}
\modulolinenumbers[5]

\usepackage{textcomp}
\usepackage[mathscr]{eucal}
\usepackage{booktabs} 
\usepackage{graphicx}
\usepackage{subcaption}
\usepackage[export]{adjustbox}
\usepackage{wrapfig}
\usepackage[T1]{fontenc}
\usepackage[ruled]{algorithm2e} 
\usepackage{multirow}
\usepackage{longtable}
\usepackage{float}
\usepackage{csquotes}
\usepackage{ltxtable}

\usepackage{fullpage}
\usepackage{times}

\usepackage{numcompress}

\begin{document}



\begin{frontmatter}

\title{Model Learning: A Survey on Foundation, Tools and Applications}

\author[mymainaddress,mysecondaryaddress]{Shahbaz Ali}

\author[mymainaddress,mysecondaryaddress]{Hailong Sun}

\author[mymainaddress,mysecondaryaddress]{Yongwang Zhao\corref{mycorrespondingauthor}}
\cortext[mycorrespondingauthor]{Corresponding author}
\ead{zhaoyw@buaa.edu.cn}

\address[mymainaddress]{School of Computer Science and Engineering, Beihang University, Beijing, China}
\address[mysecondaryaddress]{Beijing Advanced Innovation Center for Big Data and Brain Computing, Beihang University, Beijing, China}
\begin{abstract}
The quality and correct functioning of software components embedded in electronic systems are of utmost concern especially for safety and mission-critical systems. Model-based testing and formal verification techniques can be employed to enhance the reliability of software systems. Formal models form the basis and are prerequisite for the application of these techniques. An emerging and promising model learning technique can complement testing and verification techniques by providing learned models of black box systems fully automatically. 

This paper surveys one such state of the art technique called \textit{model learning} which recently has attracted much attention of researchers especially from the domains of testing and verification. This survey paper reviews and provides comparison summaries highlighting the merits and shortcomings of learning techniques, algorithms, and tools which form the basis of model learning. This paper also surveys the successful applications of model learning technique in multidisciplinary fields making it promising for testing and verification of realistic systems.
\\
\end{abstract} 


\end{frontmatter}


\section{Introduction} \label{Sec:introduction}

To imagine a software without bugs is difficult and ensuring its correctness even more difficult. The likelihood of occurring errors in systems rises with the increase of scalability and functionality. Also, the heavy use of unspecified third-party components by developers, for rapid development and reducing time-to-market, can also result in under-specified and erroneous systems. Unfortunately, bugs are common and propagated among these components. Integration testing of the overall system developed by unspecified black-box components become a challenging job. There are numerous cases where software bugs resulted in a disastrous loss of money, time, or even human life. The author \citet{dershowitz2000software} has maintained a list of over 100 \textit{"software horror stories"} including failure of Patriot missile \cite{blair1992patriot} during the Gulf war (due to software error), the crash of the Ariane 5.01 maiden flight \cite{lions1996ariane} (due to an overflow), the loss of Mars orbiter \cite{stephenson1999mars} (due to a unit error), Therac-25 radiation therapy machine \cite{leveson1993investigation} (software error), and Pentium FDIV \cite{coe1995inside} (design error in floating point division unit) are a few well-known examples highlighting the fact that life-depending, mission-critical and safety-critical systems can be far from being safe.\par

The profusion and diversity of these bugs have generated the requirements for some promising techniques to avoid, detect and correct the flaws. The versatile model-based techniques including model-based testing (MBT) and model checking are effective methods to enhance the quality and reliability of software systems by removing the bugs. The existence of behavioral models corresponding to systems under test (SUT) is the pre-requisite for the application of these techniques. These techniques are best-fit for the scenarios where behavioral models are readily available either from requirement specifications or from the use of some program analysis techniques to source code directly \cite{ball2002s,henzinger2002lazy,walkinshaw2008automated}. However, these techniques are not suitable for black-box software components which are generally distributed without access to source code and with limited documentation (or the available documentation became outdated due to system evolution). The construction of formal models is the fundamental problem for black-box software components, so we must consider some alternative approaches from the domain of software reverse engineering.\par

In recent years, a state of the art technique, i.e., \textit{Model Learning} (a.k.a automata learning) has emerged as a highly effective technique for learning the models of black-box hardware and software systems \cite{Vaandrager2017}. It complements testing and verification approaches by providing accurate and reliable models. It is a \textit{dynamic analysis} \cite{ernst2001dynamically} technique and works by posing queries to the system under learning (SUL), then infer the run-time behavioral model of a target system on the basis of received responses \cite{Vaandrager2017,Raffelt2009}. It is the test-based and counterexample-driven approach for learning models of real-world systems fully automatically.\par

Model learning approaches can mainly be categorized into active and passive learning. In passive learning, there is no interaction between the learner and SUL. The passive learning algorithms learn the behavioral models of SUL from the available set of positive and negative traces (training data) \cite{walkinshaw2007reverse,biermann1976constructing}. However, if a log file missed some behavior, then it is difficult to infer a conclusion, and hence the resulted model became erroneous due to this missed or unobserved behavior. To deal with this problem, a solution exists, and that is to request additional behavioral traces as per requirement from the target system. This strategy has been incorporated in \textit{active learning} approach. In active learning, there is a continuous interaction between the learner and SUL. The active learning algorithms learn the behavioral models of SUL by performing experiments (tests) on it. By applying this process of experimentation repeatedly, a model is approximated that represents the complete behavior of a software component. In this survey paper, we focus on \textit{active earning} approaches which provide behavioral models in the form of \textit{state diagrams}. We shall discuss \textit{passive learning} slightly for comparison purposes.\par

The learned models are highly effective and can benefit model checking approaches \cite{MllerOlm1999ModelCheckingAT,clarke1999model,baier2008principles}, model-based testing \cite{broy2005modelPap,utting2010practical}, composition of components \cite{arbab2004reo}, checking correctness of API usage \cite{ball2006thorough}, integration testing of black box components \cite{hungar2003domain,margaria2004efficient,muccini2007monitoring,shahbaz2008reverse,shu2007testing}, and to facilitate reuse \cite{LearningInOutAutomata}. Model learning techniques \cite{hagerer2002model} can also be used to check whether any new faults have been introduced in the modified version of a software component or not (regression testing) or whether an alternative implementation conforms with reference implementation or not. It is highly effective for legacy software, where the source code is not available or can hardly be understood due to the lack of documentation. Section \ref{Sec:Applications of Model Learning} presents a detailed overview of the success stories of model learning applications in multidisciplinary areas of real life.\par
 
The core part of model learning is learning algorithms and mostly follow Angluin\textquotesingle s MAT framework \cite{Angluin1987}. They usually learn the models of realistic systems in the form of DFA, Mealy machine, and register automata (EFSMs). Besides learning algorithms, testing (for validation checking) and counter-example handling algorithms have also been developed and optimized. The productive work in the development of counter-example handling algorithms has contributed a lot to reduce the learning complexity and speed up the learning process. For practical applications of model learning, we need robust and versatile implementations of learning algorithms along with surrounding infrastructure which is vital for quick assembly of learning setups. To satisfy these needs, some progressive work has been done in the development of learning libraries. Most of these libraries are free, open source and implemented in JAVA. The primary objective of these learning tools is to provide a framework for research on automata learning and their applications in real life.\par



To support our arguments about state of the art and emerging active model learning technique, we structured the rest of our paper as: next section lays out the background knowledge regarding the model. Section \ref{sec:Foundation of Model Learning}, covers the key concepts regarding the foundation of model learning. Section \ref{sec:Model Learning Tools}, highlights the progress made in the development of different model learning tools along with a comparison table. In section \ref{Sec:Applications of Model Learning}, we shall discuss the successful applications of model learning in various domains along with a summary table. Finally, the section \ref{Sec:Achievements Challenges Future Work} will cover the current achievements, challenges being faced and future trends in the field of model learning followed by the concluding remarks.


\section{Background} \label{sec:Background}
\subsection{Learning Frameworks}
The three well-known learning notions are Angluin\textquotesingle s learning model, Gold\textquotesingle s learning model, and PAC learning model. Angluin\textquotesingle s learning model is a well known formal learning model developed by Dana Angluin \cite{Angluin1987} in 1987. In this learning model, the goal of the learner is to learn the behavioral model of an unknown black box system via continuous interaction with it. This kind of learning is called \textit{query learning} (or \textit{learning via queries}) and the objective is referred as \textit{exact identification}. Angluin presented a learning algorithm $L^*$, where DFA was used to describe regular sets. The learning of DFA with L* algorithm is a typical example of polynomial-time learning via queries. It is one of the most important models of computational learning theory.

\begin{figure}[h] \centering 
\includegraphics[clip,width=0.50\textwidth,]{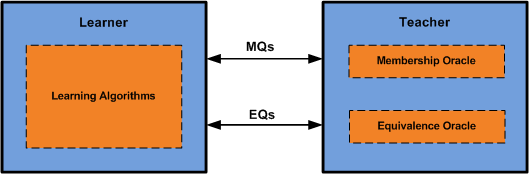}
\caption{MAT Model \cite{Merten2013ActiveAL}.}
\label{fig:The_MAT_Model}
\end{figure}

She introduced the concept of a MAT where learning process can be viewed as a game. In this game, the role of the learner is to learn an unknown model by putting the queries to a teacher/oracle. In the so-called MAT framework \cite{Angluin1987}, the construction of models involve a \textit{"learner"} and a \textit{"teacher"} as shown in figure:\ref{fig:The_MAT_Model}. The teacher knows all about the automaton SUL to be learned and on the other side learner only knows the input/output alphabets. For learning an unknown automaton $\mathscr{M}$ automatically, the learner posed queries and based on received responses it tries to construct an unknown automaton whose behavior matches with that of target automaton \cite{Cook1998,Vi}. Learner poses two types of queries:
 
\begin{itemize}
\item \textit{Membership Queries (MQs):} With \textit{MQs} the learner asks whether the input sequence belongs to the target system. The answer is \textit{"Yes"} if it belongs to, otherwise \textit{"No"}.

\item \textit{Equivalence Queries (EQs):} With \textit{EQs} the learner asks whether the hypothesized/conjectured automaton $\mathscr{H}$ is correct i.e., $\mathscr{H} \approx \mathscr{M}$. The teacher who knows about the system completely answers \textit{"Yes"} if this is the case otherwise it answers \textit{"No"} and provides a counterexample that differentiates $\mathscr{H}$ and $\mathscr{M}$ such that $\mathscr{H} \neq \mathscr{M}$.
\end{itemize} \par

Later on \citet{peled2002black} and \citet{groce2006adaptive} highlighted the fact that MAT framework can also be employed to learn the models of black box software and hardware components. A number of polynomial time algorithms utilizing MAT model have been designed and developed \cite{balcazar1997algorithms,hungar2003domain,rivest1993inference}. These algorithms were used to learn target classes such as DFA \cite{Angluin1987}, Horn sentences \cite{angluin1992learning}, read-once formulas \cite{angluin1993learning,bshouty1992learning}. Moreover, variants of the formalization of learning via queries have been proposed by \cite{watanabe1990formal,watanabe1994framework}.\par

In Gold\textquotesingle s learning model, \citet{gold1967language} introduced a learning notion in 1967 called \textit{identification in the limit or EX identification}. In order to learn this type of  learning notion there exist a framework \cite{angluin1983inductive} which has been extended \cite{gasarch1992learning} to study \textit{query learning}.  However, in this kind of learning the main issue is not the learn-ability within a given amount of time but its the learn-ability within a finite amount of time. We can say that the framework is not suitable for resource bounded-learning but for recursion theoretical learning \cite{watanabe1994framework}.\par

In probability approximately correct (PAC) learning model, for resource bounded-learning, Valiant \cite{valiant1984theory} introduced PAC learning model which is polynomial-time bounded-learning. Due to much research work in this field \cite{blumer1989learnability,pitt1988computational} there exist now a PAC learning framework \cite{pitt1990prediction}. However, the learning approach in this framework is "passive learning", which will be discussed in the next section. \citet{angluin1995won} also studied and discussed this notion of learning style in their research work.

\subsection{Model Learning Techniques}
In practice, behavioral models are often omitted because of the cost involved and to reduce time-to-market in their generation and maintenance. To address this issue, a number of different model-learning techniques were developed which infer models automatically. Depending on the utilization of source code, model learning is either white box or black box. Further, model learning approaches can be divided mainly into active and passive learning, each having its own characteristics.

\subsubsection{Passive Learning Technique}
In passive learning, there is no interaction between the learner and SUL. The passive learning algorithms, also called offline algorithms, learn the behavioral models of SUL from the available set of positive and negative traces (training data) \cite{walkinshaw2007reverse,biermann1976constructing,ammons2002mining,lo2006smartic,lorenzoli2008automatic}. Positive traces are those that belong to the target language, and negative traces do not belong to the target language. Passive learning binds the learning process to available prerecorded traces of systems that represent the behavioral aspects of the system\cite{verwer2010efficient,dallmeier2010mining}. Every trace consists of input symbol representing the stimuli the SUL was exposed to, and its output symbol representing the system\textquotesingle s reaction to the input symbol. The behavior of a system for a specific period is recorded in the form of traces in a log file. The main problem with passive learning is the quality of the learned model that depends upon the recorded traces. More recorded traces means more the availability of behavioral information to build an accurate model of the target system. Log file having missed or unobserved behavior, it is difficult to infer a conclusion, and hence the resulted model became erroneous due to this unobserved behavior.  There exists a solution to this problem, and that is to request additional behavioral traces as per requirement from the target system. This strategy has been incorporated in \textit{active learning} approach. \par

\subsubsection{Active Learning Technique}
\begin{figure}[h] \centering  
\includegraphics[clip,width=0.55\textwidth,]{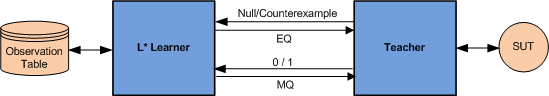}
\caption{Components of L* Algorithm}
\label{fig:Componetns of Lstar Algorithm} 
\end{figure}

In active learning, there exists continuous interaction between the learner and SUL. The active learning algorithms, also called on-line algorithms, learn the behavioral models of SUL by performing experiments (tests) on it. By applying this process of experimentation repeatedly, a model is approximated that represents the complete behavior of a software component. In the context of active learning, Dana Angluin proposed a seminal query learning algorithm \cite{Angluin1987}, $L^*$ for inferring DFA and the main components of the $L^*$ algorithm has been shown in figure: \ref{fig:Componetns of Lstar Algorithm}. Many efficient learning algorithms have been designed and developed since then, and most of them follow Angluin\textquotesingle s approach of MAT. The learner poses MQs (\textit{yes/no query}) which consist of strings from input set and record the responses 0 (which means string does not belong to language) and 1 (means string belong to language) into an observation table. In this active learning process when the table becomes \textit{close} and \textit{consistent} then learner build an automaton called conjecture/hypothesis by the help of the observation table. Now, at this point to check whether this hypothesis is equivalent to black box model (SUL), the algorithm totally depends upon the existence of an oracle. To check this validity, the learner uses EQs (\textit{yes/counterexample query}). If the black box (SUL) is not equivalent to the hypothesis, then the oracle will generate a counterexample which is a string from input set that is accepted by the black box and rejected by hypothesis and vice-versa. By analyzing the counterexample in an intelligent way the learner refines observation table and construct an improved version of the hypothesis. The whole process continues in a loop until we obtain an automaton whose behavior matches with black-box SUL.



For a regular set \textquotesingle Z\textquotesingle, the Angluin\textquotesingle s $L^*$ returns a minimal DFA $\mathscr{M}_{z}$ provided that the reply of every query is given correctly w.r.t \textquotesingle Z\textquotesingle. Moreover, $L^*$ obtains $\mathscr{M}_{z}$ in polynomial time so, we can say that regular set class, which is represented by DFA, is polynomial-time learnable. Research works \cite{angluin1988queries,berman1987learning,ishizaka1989learning,sakakibara1990learning} also have been carried out to check whether similar learnability results exist for other classes like context-free languages which are represented by context-free grammars, regular sets that can be represented by regular expressions, etc.

\subsubsection{Evaluation of Learning Techniques}
Recently, various learning algorithms have been developed for learning behavioral models of software components. It is difficult to make a decision about which algorithm is better than other. Every algorithm has its own working environment which may perform better than other algorithm in different settings. A few available benchmarks are not so diverse that can make a reliable comparison. One solution is to develop a framework which can evaluate the learning algorithms and provides an opportunity to have a look at the weaknesses and strengths of other participant\textquotesingle s algorithms. Finite state machines or grammar learning competitions like \textit{Tenjinno} that is a machine translation competition \cite{starkie2006tenjinno}, \textit{Omphalos} is a context-free grammar learning competition \cite{starkie2004omphalos}, \textit{STAMINA} \cite{walkinshaw2013stamina} and \textit{ZULU} \cite{combe2009zulu} competitions are some frameworks that can help in evaluating learning algorithms. If such competition ends with no winner, these are still useful in the sense that all the participants can have a look at the weaknesses and strengths of other\textquotesingle s learning techniques.

\section{Foundation of Model Learning} \label{sec:Foundation of Model Learning}
A system model is helpful in understanding and predicting its behavior. The behavior of the system can be characterized by input and output sequences, the flow of data with the condition from input to output actions. There are different kinds of modeling formalisms which describe various aspects of the system behavior. Examples of models include data and control flow graphs, transition-based models, dependency graphs, decision tables, stochastic models and algebraic models. State diagrams including Mealy, Moore and UML/Harel represent events, actions, and states of the system. A state model or transition based model (state-charts/FSM) can be used to express the essential behavior of the complex and realistic system. These are very useful for understanding the behavior of many practical systems like security protocols, network protocols, and embedded control softwares, etc. There are different approaches by which models of software components can be learned, for instance, by performing tests, by analysis of source code or system logs. \par

Automata learning concept has been studied in literature for decades. It has a long history and always remained a hot topic of research. In recent years, it has gained much attention in software engineering tasks, especially in testing and verification. First of all Moore \cite{moore1956gedanken} highlighted the problem of learning finite automata in 1956. Since then the problem of learning automata has been studied by different research groups under different names. Some researchers use the term of grammatical inference \cite{hungar2003domain}, some use the name of regular inference \cite{berg2005correspondence}, in literature some authors use the words of regular extrapolation \cite{hagerer2002model} or active automata learning \cite{isberner2015foundations} and security researchers selected the term of protocol state fuzzing \cite{de2015protocol}. Model learning \citeauthor{Vaandrager2017} is a more appropriate term for automata learning and it resemblance with model checking term. In model checking technology, finite-state models are analyzed whereas in model learning models of software and hardware systems are constructed by providing inputs and observing outputs. So, model learning is a complementary technique for model checking. Frits  \citeauthor{Vaandrager2017}, highlights the emergence of model learning as an extremely effective bug-finding technique and an effective method for inferring models of black-box hardware and software components. Moreover, the research work by Peled et al. \cite{peled2002black} and Steffen et al. \cite{berg2005correspondence,hagerer2002model,de2010grammatical} bring model learning and formal method techniques close to each other, particularly model checking and model-based testing.

\subsection{Basic Framework}
\begin{figure}[h] \centering  
\includegraphics[clip,width=0.50\textwidth,]{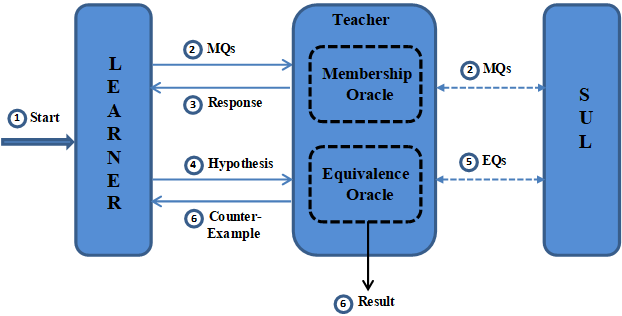}
\caption{General Learning Process with Basic MAT Framework}
\label{fig:General Learning Process}  
\end{figure}

In model learning, behavioral models for software components can be inferred using different approaches including performing tests (queries), mining of system logs (from traces) or by analysis of source code. In recent years, much progress has been made in developing efficient learning algorithms corresponding to these inferring model approaches. Among these learning approaches, active automata learning (a.k.a model learning) has gained the great attention of researchers especially from the domains of testing and verification. Most of the learning algorithms in this approach follow Angluin\textquotesingle s MAT framework (i.e., learning actively by posing queries), as shown in figure:\ref{fig:The_MAT_Model}, for learning the models of software components. A clear overview of a general learning process using MAT framework is shown in the Figure:\ref{fig:General Learning Process}.

This learning framework usually makes three assumptions for learning models. First, it assumes that the system to be learned (SUL) can be modeled as a finite automaton. Secondly, learner (learning algorithms) knows the alphabets, and it is fixed. Third, it also assumes that there is an oracle (a.k.a., teacher) who knows SUL.

\subsection{Model Learning Framework}

\subsubsection{Model Learning within MAT Framework}
Peled et al. \cite{groce2006adaptive,peled2002black} highlighted the fact that MAT framework can be utilized to infer models of hardware and software components considering them as black boxes.
The "learner" infers a model (say a Mealy machine $\mathscr{M}$) by giving inputs to SUL and observing outputs via MQs. The "learner" asks the "teacher" whether the input sequence is the part of SUL or not using MQs. Once a hypothesized model is ready, then the "learner" asks the "teacher" whether the model is correct or not via EQs. The "learner" can reset the "teacher" at any point. A reset command (a MQ) is used to bring the SUL back to its initial state. Reset is important because active learning requires MQs to be independent \cite{Steffen2011IntroductionTA}. Hence, the reset command is executed after every query. The "teacher" uses a conformance testing tool \cite{lee1996principles} to answer EQs using a finite number of testing queries (TQs). A test query (TQ) is similar to MQ and get the response of the SUL for an input sequence. If one of the TQs finds a counterexample, then it means the conjectured/hypothesized model is incorrect. In the case of the incorrect model, the answer to EQs is \textit {"No"}. Otherwise, the answer is \textit{"Yes"}, and the learning process ends. 

\begin{figure}[h] \centering  
\includegraphics[clip,width=0.40\textwidth,]{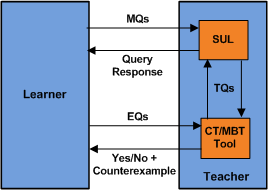}
\caption{Active automata learning in a black-box reactive system \cite{Vaandrager2017,smeenk2015applying}.}
\label{fig:Model_Learning_within_the_MAT_frameWork}
\end{figure}

A counterexample returned to the "learner" distinguishes the constructed hypothesized model and the SUL model. Based on this counter-example, the "learner" may construct an improved version of the hypothesis by giving more input sequences via MQs. A schematic overview is shown in Figure \ref{fig:Model_Learning_within_the_MAT_frameWork}. Here, the learner\textquotesingle s job is to construct hypotheses while the task of the conformance testing tool (CT) is to check the validation of the constructed hypotheses. As the testing tool (CT) can only send a finite number of TQs so we can not say with certainty that the learned model is the correct representation of the target system. However, by assuming a bound on the number of states of machine \textit{\textquotesingle M\textquotesingle}, we can have a finite and complete conformance test suite \cite{Vaandrager2017,lee1996principles}.

\subsubsection{Model Learning with a Mapper/Abstraction Component}
\begin{figure}[h] \centering  
\includegraphics[clip,width=0.55\textwidth,]{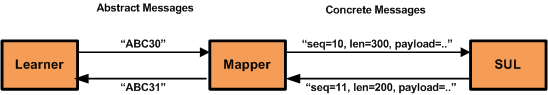}
\caption{Working of a mapper component}
\label{fig:The_Mapper_Component}
\end{figure}

Abstraction plays a crucial role in scaling existing model learning techniques to realistic systems. Mapper is a software component that plays the role of abstraction and is placed between learner and teacher. The learner sends abstract messages to the mapper component that converts them to concrete messages and forwards them to the SUL. The mapper component converts the concrete response of the SUL back to abstract messages and sends to the learner. A formal model of the abstracted interface is learned by model learning. To obtain a faithful model of SUL, the abstraction process can be reversed. Basic functioning of the mapper component is shown in Figure \ref{fig:The_Mapper_Component}. 

\begin{figure}[h] \centering  
\includegraphics[clip,width=0.40\textwidth,]{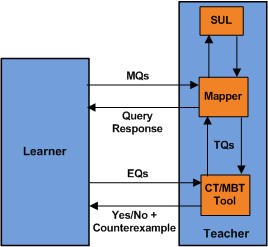}
\caption{Model Learning with mapper component \cite{Vaandrager2017}.}
\label{fig:Model_Learning_with_a_mapper}
\end{figure}

Cho et al. \cite{cho2010inference} applied model learning techniques on botnet command and control protocols for inferring models. They placed an emulator/mapper component between learning software and botnet servers. Mapper component concertizes the abstract alphabet symbols into valid network messages and forwards them to botnet servers. On receiving responses from the servers, mapper/emulator does reverse of it, i.e., it converted response messages into abstract output alphabet and forwarded them on to the learner. A schematic overview of this learning setup is shown in Figure:\ref{fig:Model_Learning_with_a_mapper}. To deal with such kind of intermediate abstraction, Aarts et al.\cite{aarts2010generating,aarts2015generating} developed a mathematical theory which incorporated the concepts of predicate abstraction and abstract interpretation. \citet{chalupar2014automated} used a different approach for inferring model. To reduce the number of inputs, they merge many input actions that have a specific order into a single high-level action \par

Researchers normally define abstraction manually. \citet{vaandrager2012active}, showed that such an abstraction component could be created automatically for the richer class of models, e.g., EFSMs. In such modeling formalism, equality of data parameters can be tested, but operations on data are not allowed \cite{Aarts2012AutomataLT,Aarts2012ATO}. Their approach makes use of counterexample-guided abstraction refinement (CEGAR) notion. In CEGAR concept, whenever the current abstraction becomes very coarse and causes non-determinism, the employed abstraction is refined automatically. In his talk, \citet{vaandrager2012active} compared his technique with the work done by Howar et al \cite{Howar2012InferringCR,Merten2012DemonstratingLO} on register automata model.
 
\subsection{Learned Models} \label{sec:Learned Models}
Having a formal representation of a system is prerequisite for verification techniques like model checking and model-based testing. Different modeling formalisms exist, including DFA, NFA, Mealy machine, Moore machine, and variants of EFSMs,  to model systems formally. We shall briefly discuss the following well-known models that are being used in model learning for realistic systems.

\subsubsection{Deterministic Finite Automaton}
A well-known modeling object in language theory is DFA that model and recognize regular languages. Some efficient learning algorithms for minimized automata models resulting in canonical forms \cite{merten2013active,Angluin1987} have been proposed (see table:\ref{table:Comparison of Different Learning Algorithms}). Automata learner for DFA (e.g., Angluin\textquotesingle s L* algorithm) learns the DFA that represents the behavior of a target automaton.

A DFA is a tuple \textit{$ M=(Q,\ q_{0},\ I,\ \delta,\ F) $} where (1) $ Q $ is finite non empty set of \textit{states} (2) $ q_{0}\ \epsilon\ Q $ is the \textit{initial} state (3) $ I $ is Input \textit{alphabet} set (4) $ \delta:\ Q\ \times \ I\rightarrow Q $ is the \textit{transition function} (5) $ F\ \subseteq\ Q $ is the finite set of \textit{accepting or final} states.



\begin{figure}[h] \centering  
\includegraphics[clip,width=0.55\textwidth,]{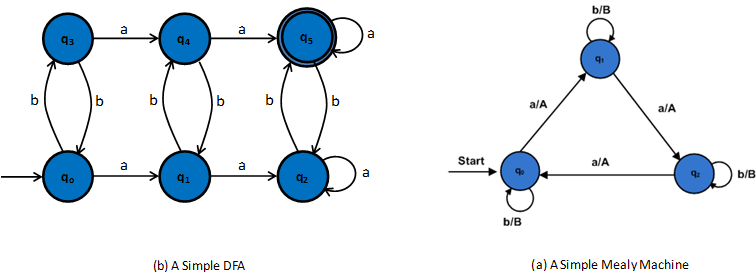}
\caption{A Simple Mealy Machine}
\label{fig:Basic Learning Models} 
\end{figure}

Intuitively, a DFA evolves as: let initially it is in some state $ q\ \epsilon\ Q $, and upon receiving an input symbol (or action) $ i\ \epsilon\ I $, it transfers to a new state according to $ \delta\ (q,i)$. A word $ i\ \epsilon\ I^{*} $ is accepted by the finite state machine if we start from the initial state and the DFA reaches to an accepting state by traversing the symbols of the word. Figure:\ref{fig:Basic Learning Models}(a) represents a simple example of a DFA that accepts a regular language over the alphabet ${a,b}$. This DFA accepts all strings having an odd number of b\textquotesingle s and at least two a\textquotesingle s.

\subsubsection{Mealy Machine}  \label{sec:Mealy machine}
In reality, reactive systems are more common than DFAs. Also, \textit{DFAs} are not suitable candidates for modeling reactive systems because of their limited expressiveness. To model reactive systems(input/output based systems) \textit{Mealy machine} is better option. These systems interact with their environment and work in a cyclic loop. They take inputs, process them and return outputs \cite{M2013Sindhu}. Various algorithms which based on Angluin\textquotesingle s approach and model the systems in the form of a Mealy machines have been proposed \cite{niese2003integrated,steffen2011introduction,margaria2004efficient} (see table:\ref{table:Comparison of Different Learning Algorithms}). A slightly different approach to model systems in the form of Mealy machine is suggested by \cite{shahbaz2009inferring,shu2007testing} that replaces the membership queries by output queries and record the output responses directly into the observation table.



 





Formally a (deterministic) mealy machine is a tuple \textit{$ M\ =\ (I,\ O,\ Q,\ q_{0},\ \delta,\ \lambda) \ $} where (1) \textit{I} is a set of input symbols/actions (\textit{the input alphabet}) (2) O is a set of output symbols (\textit{the output alphabet}) (3) Q is a finite set of \textit{states}, also called \textit{locations} (4) $ q_{0}\ \epsilon\ Q $ is the \textit{initial} state (5) $ \lambda:\ Q\ \times\ I\rightarrow O $ is \textit{output function}. It produces an output symbol for every transition (6) And $ \delta$ is the transition function which is defined as: $ \delta:\ Q\ \times\ I\rightarrow Q $. It specify the transitions for every state to its successor states. Figure \ref{fig:Basic Learning Models}(b): represents a simple Mealy machine. Intuitively, it evolves as: let initially it is in some state $ s\ \epsilon\ S $, and whenever it receives an input symbol $ a\ \epsilon\ \Sigma $, the machine change its state to a new one according to $ \delta (s,a) $. Its functioning is very similar to a DFA but in addition it also generates an output symbol $ o\ \epsilon\ O $ according to $ \lambda(s,a) $.


\subsubsection{EFSM / Register Automaton} \label{sec:EFSM and RA}
\begin{figure}[h] \centering  
\includegraphics[clip,width=0.55\textwidth,]{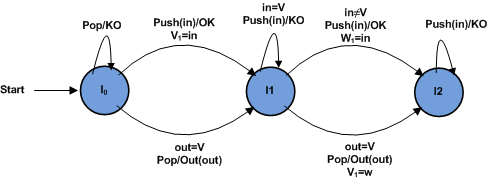}
\caption{A Register Automaton \cite{vaandrager2012active}.}
\label{fig:A_Register_Automaton}
\end{figure}

Mealy machines represent the control flow of the target component, and it is also limited in its power of expressiveness. It becomes difficult to model realistic systems like software protocols with the Mealy machine. In this case, we use extended finite state machine (EFSM) and its variants like register automata (RA). EFSMs can model data flow as well as control flow of software module since input and output symbols can carry data values. A common EFSM formalism is the register automata, in which finite control structure is combined with variables, guards and assignments. Some work has been done to generalize these learning algorithms to richer models like EFSMs where data values can be communicated, processed and stored \cite{howar2012inferring,merten2012demonstrating,isberner2014learning,howar2012inferringSemantic}. Figure:\ref{fig:A_Register_Automaton} represents a register automaton model for a FIFO-set with capacity two. Interested readers are referred to \cite{vaandrager2012active,aarts2014algorithms} for further details.


\subsubsection{Miscellaneous Formalisms} \label{sec:miscellaneous formalisms}
Some other modeling formalisms for which learning algorithms have been developed include: Nondeterministic auotmata \cite{yokomori1994learning,denis2000learning}, Probabilistic automata \cite{clark2004pac,castro2008towards}, Petri-nets \cite{van2011process}, Timed automata (learning timed systems in active learning framework) \cite{grinchtein2010learning,verwer2012efficiently}, Buchi automata \cite{de2004inference} and I/O automata \cite{LearningInOutAutomata}.

\subsection{Data Structures}
Model learning algorithms normally vary in two ways: (1) the data structures used for storing responses of queries and realizing the black-box abstraction, (2) and how learning algorithms handle counterexamples. The basic learning algorithm L* and its variants used two types of data structures: \textit{observation tables} and \textit{discrimination trees}. 

\subsubsection{Observation Tables}
\begin{figure}[h] \centering  
\includegraphics[clip,width=0.55\textwidth,]{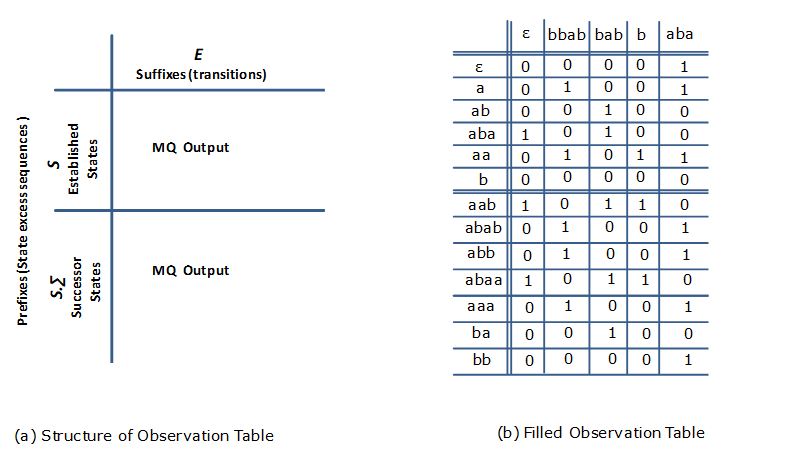}
\caption{The structure of L* observation table.}
\label{fig:Observation Table}
\end{figure}

Observation table is the most well-known data structure, and it was introduced first by Gold \cite{gold1978complexity}. Later on, Angluin used this data structure in her seminal algorithm L* (learner) for organizing the information collected during the interaction with the system under learning (SUL). Different learning algorithms have used variants of this observation table for other modeling formalisms. The table may be a two-dimensional array that is characterized by upper and lower parts. The data values in the upper part are used to represent the states of minimal state model, and the lower part is used for transitions. Automaton\textquotesingle s states can be distinguished by sets of words named as "prefixes" and "suffixes". We can denote the observation table as a triple over an alphabet $\Sigma$ by $(S,E,T)$. The structure of an observation table has been shown in the figure:\ref{fig:Observation Table} (a). Here, S and E are sets of strings over $\Sigma$ which are non-empty and finite. Also, $S\subseteq\Sigma^{*}$ is \textit{prefix closed} and $E\subseteq\Sigma^{*}$ is \textit{suffix closed} sets, and T is a finite function which is defined (for Angluin\textquotesingle s Lerner L*) as $T\ : \ ((SUS.\Sigma)\ x\ E)\ \rightarrow\ \lbrace 0,1\rbrace$. The rows and columns are labeled with $(SUS.\Sigma)$ and E respectively. For any row $s\ \epsilon \ (SUS.\Sigma)$ and column $e\ \epsilon \ E$, then $T(s,e)$ will represent the corresponding cell in table. The value of T(s,e) is "1" if the string s.e is accepted by target SUL and "0" otherwise, as shown in the figure:\ref{fig:Observation Table} (b).  

For other modeling formalism like a Mealy machine (for example), \citet{shahbaz2009inferring} slightly modified the structure of observation table. They recorded the behaviors of the system as strings in the table instead of \enquote{0} and \enquote{1}. When the above table becomes \textit{"close"} and \textit{"consistent"} then a hypothesis is constructed. The rows labeled with set S form the states of model and columns labeled with set E distinguish these states. The elements of the lower part, i.e., $S.\Sigma$ are used in building transitions.\par

An observation in observation table can be denoted by a tuple $(In_{str},Out_{str})\ \epsilon \ (\Sigma^{*}\ *\ O^*)$ for which $|In_{str}|=|Out_{str}|$. A Mealy machine  $\mathscr{M}$ having start state $q_{0}$ holds a set of observations $obs_{\mathscr{M}}$ defined by: $obs_{\mathscr{M}} = \left\lbrace (In_{str},Out_{str}) | \lambda(q_{0},In_{str})=Out_{str} \right\rbrace$. If we consider the Mealy machine in Figure: \ref{fig:Basic Learning Models}(b), then for input string $In_{str}=abab$ and starting state $q_{0}$ the machine will return output as $Out_{str}=ABAB$. So, $(abab,ABAB)\ \epsilon \ obs_{\mathscr{M}_{example}}$. 


\subsubsection{Discrimination Tree}
The second data structure used by many learning algorithms is a discrimination tree which is a decision tree for determining equivalence of states. It is a classification scheme for distinguishing between states and first introduced by Kearns \& Vazirani \cite{kearns1994introduction}.

\begin{figure}[h] \centering  
\includegraphics[clip,width=0.30\textwidth,]{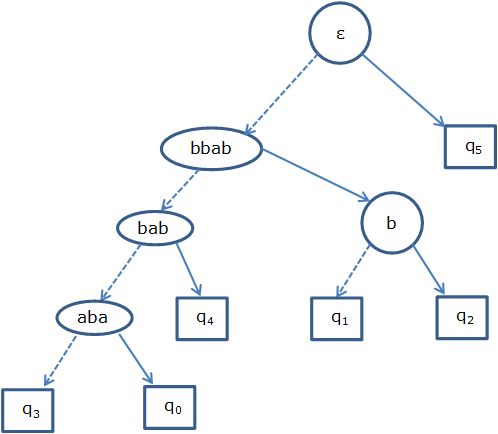}
\caption{Discrimination Tree for SUL in Figure:\ref{fig:Basic Learning Models}(a), using \textit{Observation Pack} algo \cite{Isberner2015LearnLibT}.}
\label{fig:Discrimination Tree} 
\end{figure} 

Figure: \ref{fig:Discrimination Tree} represents a discrimination tree that is obtained by applying Observation Pack learning algorithm on SUL having formal behavior shown in figure:\ref{fig:Basic Learning Models}(a). Here, leaves represent the states of hypothesis and discriminators labeled the inner nodes. Further, every node holds two children, i.e., 0-child (represented by dashed line) and 1-child (represented by solid line). 
Two well-known operations which are used to get information from a discrimination tree are: (1) \textit{"Sifting"} and (2) computing \textit{"lowest common ancestor (LCA)"}. The "sifting" explores the tree for classification, and "LCA" emphasizes the separation of classes present in a tree. For further details about discrimination tree data structure, we refer the interested readers to \cite{Isberner2015LearnLibT,Isberner2015FoundationsOA}.

\subsection{Learning and Testing Algorithms}

\subsubsection{Learning Algorithms} Learning algorithms are the backbone of automata learning techniques for obtaining models of realistic systems. These algorithms are also playing a key role in various domains of computer science including AI, data mining, neural networks, robotics, geometry, pattern recognition, natural language processing and particularly in verification and testing. Many learning algorithms have been designed and developed, and most of them follow Angluin\textquotesingle s MAT framework. These algorithms can be differentiated on the basis of data structures, modeling formalism and the learning approach (active or passive) \cite{aarts2014tomte} they used. Some well-known algorithms are L* \cite{Angluin1987}, NL* \cite{bollig2009angluin}, ADT, DHC \cite{merten2012automata}, Maler\&Pnueli\cite{maler1995learnability}, Shahbaz \& Groz \cite{shahbaz2009inferring}, Suffix1by1 \cite{irfan2013model}. Kearns and Vazirani \cite{kearns1994introduction} made an improvement in $L^*$ algorithm. They replaced the observation structure with a discrimination tree and developed a new learning algorithm.  Rivest and Schapire \cite{rivest1993inference} also noted an inefficiency in $L^*$ algorithm regarding counterexample\textquotesingle s prefixes which were being added as rows in the observation table. They pointed out that it is unnecessary to add all the prefixes of a counterexample to observation table as rows. This discrepancy can be removed by adding a single and well-selected suffix to table as a column. TTT algorithm designed and developed by Isberner et al. \cite{isberner2014ttt,isberner2015foundations} is believed to be the best algorithm among the available model learning algorithms \cite{Vaandrager2017}. This algorithm was designed and developed by considering the inefficiencies in the algorithms proposed by Kearns \& Vazirani \cite{kearns1994introduction} and Rivest \& Schapire \cite{rivest1993inference}. TTT algorithm efficiently eliminates the excessively long discrimination trees, which may be resulted while processing long counterexamples. A few well-known passive learning algorithms include RPNI, RPNI-MDL, RPNI-EDSM, DeLeTe2, and OSTIA\_4MM. Table: \ref{table:Comparison of Different Learning Algorithms} gives a summary of these algorithms using different characteristics like learning approach, modeling formalism, implementing library, etc.


 
\subsubsection{Testing Algorithms} \label{sec_testing algorithms} Algorithms that generate test suites are of great importance and playing their crucial role in automata learning. Once a learning algorithm constructs a hypothesis then testing algorithm act as an \textit{oracle (equivalence oracle)} to check whether the hypothesis is correct or not. In general, there are two kinds of oracles: \textit{ideal oracle} and \textit{real oracle}. The ideal oracle knows the underlying automaton of SUL, and in case of real oracle it does not have access to the automaton of SUL, and it has to approximate the equivalence queries by testing. The real oracle is connected to a system, and it has to approximate it. In case of software system\textquotesingle s implementations, \textit{Real oracle} does not exist. We can alleviate this problem by using \textit{"random sampling oracle"}. Random sampling explores the learned model and SUL to search for discrepancies. A wide variety of algorithms (which act as oracles) have been proposed for different classes of models. Automata learning libraries like LearnLib provides testing algorithms besides learning algorithms. These algorithms include state cover set, transition cover set, Random words, W-Method \cite{chow1978testing}, Random walk, Wp-Method \cite{fujiwara1991test}, W-method randomized, Wp-method randomized, UIO method \cite{shen1992protocol} and UIOv method \cite{vuong1990uiov}. 
 
\subsection{Learning Complexity}
Learning algorithm\textquotesingle s complexity is normally calculated in terms of the required number of membership and equivalence queries. It is due to the fact that the execution of membership queries requires interaction with the SUL and it will take some time. The observation table or other data structures like reduced observation table or discrimination tree for queries and responses need to allocate memory resources. Let the size of the alphabet $\Sigma$, is denoted by $|\Sigma|$ , $\textit{\textquotesingle n\textquotesingle}$ be the total number of states in the target minimal state model of SUL, size of input alphabet \textit{I} is \textit{|I|}, and $\textit{\textquotesingle m\textquotesingle}$ be the longest counterexample returned by the oracle. Let we first discuss the complexity of equivalence queries. In Angluin\textquotesingle s algorithm, and the algorithms that are using other structures like discrimination tree or reduced observation table, the upper bound for the required number of equivalence queries is $\textit{\textquotesingle n\textquotesingle}$. \par

Now we discuss the upper bounds for membership queries which are not as simple as the case of equivalence queries. The upper bound on algorithms like Angluin\textquotesingle s that uses \textit{observation table} data structure is $O(|\Sigma|mn^{2})$ \cite{shahbaz2009inferring}. The algorithms like Rivest \& Schapire \cite{rivest1993inference} that use \textit{reduced observation table} data structure  (a reduced version that stores a smaller portion of queries and their responses) has upper bound on membership queries as $O(|\Sigma|n^{2}+nm)$. The third type of algorithms like Kearns \& Vazirani \cite{kearns1994introduction} use \textit{binary discrimination tree}, a completely different data structure for information storing. Such category of algorithms has upper bound on membership queries as $O(|\Sigma|n^{2}+n\log m)$. The upper bounds on membership queries depend on responses to these queries whether they are stored or not (not saved in case of discrimination tree but saved on other two cases) and how many these membership queries are posed before creating hypothesis \cite{bohlin2008regular,berg2006regular}. Among these three categories of algorithms, Angluin\textquotesingle s algorithm normally poses more queries to build a model and hence it collects more information. Due to this reason, Angluin\textquotesingle s algorithm produces less false hypotheses and thus fewer equivalence queries. For these three categories of algorithms, the upper bound for equivalence queries is however same, i.e., \textit{n}.\par

The complexity of Angluin\textquotesingle s and its variants mainly depends upon the size of the alphabet, the length of counterexample and number of states. In the learning process the required number of MQs increase linearly with the number of inputs, with the length of counterexample and quadratically with the number of states \cite{isberner2015foundations}. \citet{berg2005insights} analyzed the performance of Angluin\textquotesingle s algorithm by considering randomly generated automata and real-world examples. They studied the impact of alphabet size $|\Sigma|$ and the total number of states $\textit{n}$ on the required number of membership queries (MQs). As complexity also depends upon the length of counterexample, so handling counterexamples efficiently, a number of algorithms have been proposed \cite{rivest1993inference,maler1995learnability,shahbaz2009inferring,shahbaz2008reverse,irfan2010state}. \par 

During the learning process, the formation of a hypothesis is a somewhat easy task, but its validation using conformance testing becomes a challenging task for large input alphabet. Let, n be the number of states present in learned hypothesis \textit{(H)}, $\acute{n}$ be the states present in system under learning (SUL), and if $\textit{\textquotesingle p\textquotesingle}$ be number of inputs then in worst scenario, we require to run test sequences of $\acute{n}-n$ inputs that is, $p^{(\acute{n} - n)}$ possibilities \cite{lee1996principles,Vaandrager2017}. So, there is a requirement for some strategies that can reduce the number of inputs. One solution is to use \textit{abstraction} technique.

\subsection{Comparison}
   
\begin{center}

\begin{longtable}{|p{2.7cm}|p{1.1cm}|p{1.5cm}|p{1.7cm}|p{1.8cm}|p{4.4cm}|p{1.6cm}|}

\caption{Comparison of Different Learning Algorithms}
\label{table:Comparison of Different Learning Algorithms}\\
\hline

\textbf{ALGORITHMS \#}  &  \textbf{Learn-\newline ing\newline Type} & \textbf{Data\newline Structure} & \textbf{Basic Learning Models}\newline (DFA,NFA,\newline FSM) & \textbf{Richer Learning Models}\newline (VPDA, EFSM/\newline PFSM/RA) & \textbf{Key Features}& \textbf{Reference}\\ \hline  
\endfirsthead

\multicolumn{3}{c}%
{{\bfseries \tablename\ \thetable{} -- continued from previous page}} \\ 
\hline 

\textbf{ALGORITHMS \#}  &  \textbf{Learn-\newline ing\newline Type} & \textbf{Data\newline Structure} & \textbf{Basic Learning Models}\newline (DFA,NFA,\newline FSM) & \textbf{Richer Learning Models}\newline (VPDA, EFSM/\newline PFSM/RA) & \textbf{Key Features}& \textbf{Reference}\\ \hline 
\endhead
 
\hline \multicolumn{7}{|r|}{{Continued on next page}} \\ \hline
\endfoot

\hline
\endlastfoot

\hline 
\textbf{L*}&Active&Observation Table&DFA&No&\textsl{Implementation}: LearnLib; Libalf; AIDE and RALT\newline \textsl{Complexity}: $O(|\Sigma|mn^{2})$&\cite{Angluin1987}\\
\hline 
\textbf{$L^{*}_{col}$}\newline(\enquote{Maler/Pnueli})&Active&Observation Table&DFA&No&\textsl{Implementation}: Libalf; \newline \textsl{Complexity}: $O(|\Sigma|mn^{2})$&\cite{Maler1991OnTL}\\
\hline
\textbf{Mealy Machine algo}\newline $L_M$ (adaption of L*) &Active&Observation Table&Mealy&No&\textsl{Implementation}: RALT\newline \textsl{Complexity}: $O(|\Sigma|mn^{2})$&\cite{shahbaz2009inferring}\\
\hline  
\textbf{Mealy Machine algo} \newline $L_M+$ (CE handling improvements) &Active&Observation Table&Mealy&No&\textsl{Implementation}: RALT\newline \textsl{Complexity}: $O(|\Sigma|mn^{2})$&\cite{Irfan2010AngluinSF,shahbaz2009inferring}\\
\hline  
\textbf{ADT}\newline (Adaptive Discrimination Tree) &Active&Discrimina-tion Tree&FSM (Mealy)&No&\textsl{Implementation}: LearnLib& --\\
\hline 
\textbf{DHC}\newline (Direct Hypothesis Construction)&Active&-&Mealy&No& \textsl{Implementation}: LearnLib; AIDE;\newline
\textsl{Data Structure}: Direct construction of hypothesis from observations, i.e., without observation table. Suited in the environment where memory is critical issue.&\cite{Merten2011AutomataLW} \\ 
\hline
\textbf{Discrimination Tree}&Active&Discrimina-tion Tree&DFA, Mealy&VPDA&\textsl{Implementation}: LearnLib\newline \textsl{Complexity}:  $O(|\Sigma|n^{2}+n\log m)$&\cite{kearns1994introduction}\\
\hline
\textbf{Kearns\textendash Vazirani}\newline(original)&Active&Discrimina-tion Tree&DFA, Mealy&No&\textsl{Implementation}: LearnLib; Libalf\newline \textsl{Complexity}:  $O(|\Sigma|n^{2}+n^{2}m)$&\cite{kearns1994introduction}\\
\hline  
\textbf{Kearns\textendash Vazirani}\newline(bin search)&Active&Discrimina-tion Tree&DFA, Mealy&No&\textsl{Implementation}: LearnLib \newline \textsl{Complexity}:  $O(|\Sigma|n^{2}+n^{2}\log m)$&\cite{Isberner2014AnAF}\\
\hline  
\textbf{PFSM algorithm $L_P*$}&Active&Observation Table&No&PFSM&\textsl{Implementation}: RALT\newline \textsl{Complexity}: $O(|\Sigma|mn^{2})$&\cite{li2006integration}\\
\hline 
\textbf{$L_{1}$}/Suffix1by1\newline &Active&Observation Table&Mealy&No&\textsl{Implementation}: AIDE\newline \textsl{Complexity}: $O(|\Sigma|mn^{2})$&\cite{irfan2012analysis}\\ 
\hline 
\textbf{Observation Pack}&Active&Discrimina-tion Tree&Mealy&No&\textsl{Implementation}: AIDE; It Builds upon Rivest\&Schapire\textquotesingle s algorithm; \newline \textsl{Complexity}: $O(|\Sigma|n^{2}+n\log m)$&\cite{FalkHowarPhDTheis,Isberner2014AnAF}\\
\hline
\textbf{NL*}&Active&Observation Table&NFA&No&\textsl{Implementation}: LearnLib; Libalf. Learning algorithm learns NFA using MQs and EQs. More specifically,residual finite-state automata (RFSA) are learned&\cite{bollig2009angluin}\\
\hline
\textbf{TTT}&Active&Discrimina-tion Tree&DFA, Mealy&VPDA&\textsl{Implementation}: LearnLib. It mitigate the effects of long-counterexamples; \newline \textsl{Complexity}:  $O(|\Sigma|n^{2}+n\log m)$&\cite{Isberner2015LearnLibT}\\
\hline
\textbf{Visibly 1-counter automata}&Active&Observation Table&DFA&No&\textsl{Implementation}: Libalf&\cite{neider2010learning}\\
\hline
\textbf{K-Equivalent Method}&Active&Observation Table&Mealy&No&\textsl{Implementation}: RALT&\cite{groz2008modular}\\
\hline
\textbf{RPNI}&Passive&Log of pre-recorded traces&DFA, Mealy&No&\textsl{Implementation}: LearnLib; Libalf and AIDE&\cite{Daelemans2010ColinDL}\\
\hline
\textbf{RPNI\textendash EDSM}&Passive&Log file&DFA&No&\textsl{Implementation}: LearnLib&\cite{Daelemans2010ColinDL}\\
\hline
\textbf{RPNI\textendash MDL}&Passive&Log file of traces&DFA&No&\textsl{Implementation}: LearnLib&\cite{Daelemans2010ColinDL}\\
\hline
\textbf{DeLeTe2}&Passive&Log file&NFA&No&\textsl{Implementation}: Libalf&\cite{Abela2004GrammaticalIA}\\
\hline
\textbf{Biermann \& Feldman\textquotesingle s}&Passive&Log of pre-recorded traces&NFA&No&\textsl{Implementation}: Libalf&\cite{Abela2004GrammaticalIA}\\
\hline
\textbf{Biermann \& Feldman\textquotesingle s}\newline algorithm (using SAT-solving) &Passive&log file&DFA&No&\textsl{Implementation}: Libalf&\cite{Abela2004GrammaticalIA}\\
\hline
\textbf{$OSTIA\_4MM$}&Passive&log file of traces&Mealy&No&\textsl{Implementation}: AIDE&\cite{Abela2004GrammaticalIA}\\

\end{longtable}
\end{center}

\subsection{Benchmarks}

Various benchmarks have been used for the evaluation of learning and conformance testing algorithms. Randomly generated machines (as benchmarks) were also used to evaluate testing techniques, e.g., research work by \citet{sidhu1989formal}, \citet{dorofeeva2005experimental}, \citet{bollig2009angluin} and by \citet{berg2005insights}. However, these benchmarks are small, academic or randomly generated which do not properly evaluate efficiency. Also, some experimental work proved this fact that the performance of these algorithms on such kind of benchmarks is often different from the performance on models of real systems that occur in practice.\par

The Radboud University has established a Wiki\footnote{http://automata.cs.ru.nl} page and shared publicly a number of well tested and useful benchmarks\footnote{http://automata.cs.ru.nl/Overview} of state machines that model real protocols and embedded systems. Researchers can use these benchmarks for comparing the performance of learning and testing algorithms. \citet{aarts2014algorithms} in his research work compared two established approaches of inferring register automata. The authors set up a repository\footnote{https://github.com/LearnLib/raxm} of benchmarks regarding register automaton models received from a variety of sources and application domains. They range from explanatory examples that have been discussed in the literature, to manually written specifications of data structures, to models that were inferred from actual systems (e.g., the Biometric Passport and the SIP protocol).\par

Models provided by Edinburgh concurrency workbench (CWB) can also be used as benchmarks to evaluate learning and conformance testing algorithms. CWB \cite{stevens1999edinburgh} is a tool that is used for analysis and verification of concurrent systems. This tool provides a number of fabricated finite models for realistic systems including vending machine, mailing system, ABP and ATM protocols, etc. All synthetic models provided by CWB have a different number of states and input set size.

\section{Model Learning Tools} \label{sec:Model Learning Tools}

\subsection{Tools} \label{sec:Tools}
To make model learning practical, we needed effective and efficient implementations of learning algorithms. Besides algorithms, surrounding infrastructure is also compulsory for rapid assembling of learning setups. To satisfy these needs for a flexible and comprehensive framework for model learning, different learning libraries have been developed. Followings are some well known learning libraries/tools that are being used widely in the domain of model learning.

\subsubsection{AutomataLib}
AutomataLib\footnote{https://learnlib.de/projects/automatalib} library is free, open source and implemented in Java. It supports modeling automata, graphs, and transition systems. It was developed at the Dortmund University of Technology, Germany. Its main objective is to serve LearnLib, another Java library which will be discussed in next section, However, the implementation of AutomataLib is completely  independent of LearnLib and can be used for other projects as well. AutomataLib supports some selected classes of NFA, but it mainly focuses on DFA. The current version of AutomataLib (0.7.1) supports modeling of generic transition systems, deterministic finite automata , Mealy machines and advanced structures such as visibly pushdown automata. Support for  other structures like EFSMs and its variant like register automata (RA) will be incorporated in future updates.\par

\begin{figure}[h] \centering  
\includegraphics[clip,width=0.55\textwidth,]{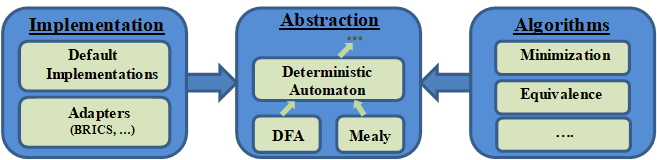}
\caption{Architecture of AutomataLib \cite{isberner2015open}.}
\label{fig:Architecture of AutomataLib} 
\end{figure}

AutomataLib is mainly divided into three parts namely implementation, abstraction and algorithms as shown in figure:\ref{fig:Architecture of AutomataLib}. Implementation part contains generic automaton implementations of DFAs and Mealy machines. Different Adapters, e.g., adapter for BRICS library is also contained an implementation part. The abstraction layer consists of a number of Java interfaces for the representation of various automata types. Algorithms part consists of algorithms for minimization, equivalence or visualization. Dot tool from GraphVIZ\footnote{https://www.graphviz.org} can be used for visualization. Interested readers are referred to \cite{isberner2015open} for further detail.

\subsubsection{Learnlib}
\begin{figure}[h] \centering  
\includegraphics[clip,width=0.60\textwidth,]{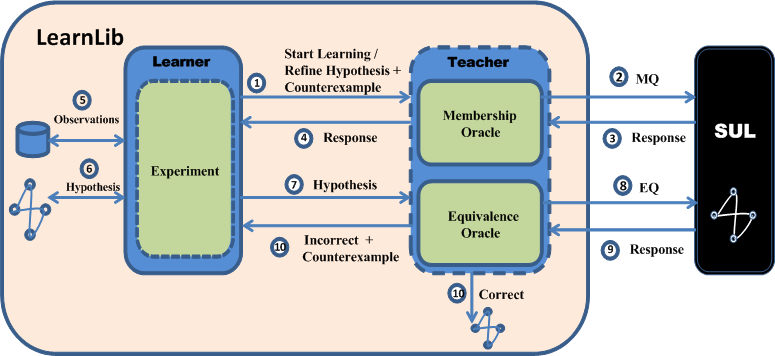}
\caption{Overview of learning model with LearnLib}
\label{fig:Learning a model with LearnLib}
\end{figure}

LearnLib\footnote{https://learnlib.de/} library was developed at the Dortmund University of Technology, Germany. The library is free, open source and implemented in Java for automata learning algorithms. The sole objective of LearnLib is the provision of a framework for conducting research work on automata learning algorithms and the applications of these algorithms in real life. It provides a number of learning algorithms for active as well as passive learnings corresponding to different modeling formalisms. It also provides different equivalence checking algorithms. Its modular and generic architecture allows us to extend it easily. The user can start learning experiments that are modified according to their needs \cite{Steffen2011, 10.1007/978-3-319-21690-4_32}. It provides the facility to analyze the learning  statistics including no of membership \& equivalence queries, runtime, and memory consumption \cite{Raffelt2005,isberner2015open}. In recent years, it is being used successfully to learn the behavioral models of realistic systems. Figure:\ref{fig:Learning a model with LearnLib} shows the schematic overview of learning a model with LearnLib library.\par

LearnLib is consist of three main modules, as shown in Figure \ref{fig:LearnLib_Basic_Components}, which are automata learning module, infrastructure module, and equivalence queries module.
  
\begin{enumerate}

\item \textsl{Automata Learning Module: }  
This is the main module of LearnLib that comprises different learning algorithms and their supported modeling structures. It also provides algorithms for handling data structures efficiently which enable learning techniques to learn large-scale systems \cite{isberner2015open}. Besides, as the way of processing counterexamples has a significant impact on the learning complexity, so Learnlib has been equipped with efficient counterexample handlers. Table\ref{table:Comparison of Different Learning Algorithms} shows the details of these learning algorithms along with other features.

\begin{figure}[h] \centering  
\includegraphics[clip,width=0.55\textwidth,]{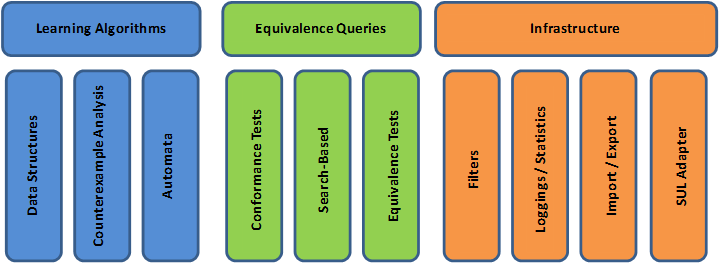}
\caption{An overview of basic LearnLib components. \cite{learnLibWebSite}}
\label{fig:LearnLib_Basic_Components}
\end{figure}




\item \textsl{Infrastructure Module: }
This module provides the infrastructure for query optimization, utilities for statistical parameters (e.g., no of MQs, no of EQs, memory consumption, running time, etc.), logging facility, a mechanism for storing and loading of hypotheses, and facility of SUL adapter. One of the goals of this module is the reduction of MQs so that this technique can be applied to realistic systems. To achieve this goal, this module provides a number of filter techniques that make use of domain and expert knowledge. To eliminate duplicate queries posed by learning algorithms, \textit{cache filters} \cite{margaria2005knowledge} were used. Parallelization component of this module divide queries among multiple workers \cite{howar2012teachers}. Learning algorithms in combination with filters can be employed to optimize queries \cite{Hungar2003}. The figure:\ref{fig:Interplay of basic LearnLib components} shows how the components of this module play their role with other modules in the learning process.

\begin{figure}[h] \centering  
\includegraphics[clip,width=0.55\textwidth,]{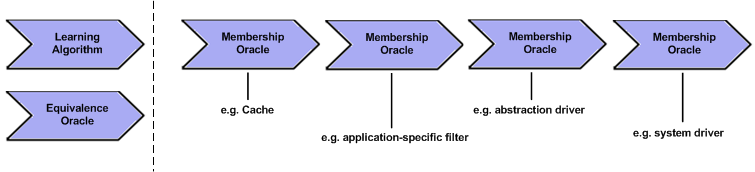}
\caption{Interplay of basic LearnLib components \cite{Merten2013ActiveAL}.}
\label{fig:Interplay of basic LearnLib components}
\end{figure}

\item \textsl{Equivalence Queries Module: }
This module depends upon search-based techniques, conformance testing, and equivalence tests \cite{Raffelt2006,isberner2015open,10.1007/978-3-319-21690-4_32,10.1007/978-3-642-38088-4_9}. The process of finding a counterexample can be seen as an \textit{equivalence query}. For a known target system, "perfect" equivalence queries are possible, and LearnLib uses Hopcroft and Karp's algorithm\cite{almeida2009testing,hopcroft1971linear} implemented in AutomataLib. In the case of the black-box target system, equivalence queries can be approximated, and queries can be answered by testing the target systems \cite{hagerer2002model,steffen2003behavior}. AutomataLib provides implementations of conformance tests that can find missing states. Some of these implementations are W-method \cite{chow1978testing}, Wp-Method \cite{fujiwara1991test}, W-method randomized, Wp-Method randomized, random words, random walk, and complete depth bounded exploration. 
\end{enumerate}

\subsubsection{LearnLib-based Developed Tools}
It was used in the development of tools like PSYCO \cite{giannakopoulou2012symbolic,howar2013hybrid}, Tomte \cite{aarts2012automata} and ALEX \cite{bainczyk2016alex}. In white-box scenarios (where source code is accessible), the Psyco and Sigma* \cite{botinvcan2013sigma} tools combine L* algorithm with symbolic execution to infer behavior models. Tomte\footnote{http://tomte.cs.ru.nl/}, which was developed at Radboud University of Nijmegen, utilize learning algorithms provided by LearnLib for learning richer classes of models (e.g., EFSMs). Tomte fully automatically constructs abstractions for automata learning. Automata Learning Experience (ALEX\footnote{https://learnlib.de/projects/alex/}) is an easy to use tool which is developed in JAVA to infer automaton models of web applications and JSON-based web services, based on top of LearnLib. LearnLib has the functionality to provide a hypothesis in the form of a Mealy machine which is represented in the standard graph description language like GraphViz. 

\subsubsection{The Libalf}
The Libalf\footnote{http://libalf.informatik.rwth-aachen.de/} is an \textit{Automaton Learning Framework} and developed at RWTH Aachen. It is free, open-source and a comprehensive library for learning finite-state automata \cite{bollig2010libalf}. Libalf implemented some well-known learning techniques such as Angluin\textquotesingle s L*, RPNI, and Biermann\textquotesingle s learning approach. Libalf is highly flexible, modular in design and learning algorithms targetting other formalisms like timed automata, Buchi automata, or probabilistic automata can also be incorporated in it. The supported algorithms along with corresponding target models are summarized in the Table:\ref{table:Comparison of Different Learning Algorithms}.





\subsubsection{Automata-Identification Engine (AIDE)}
AIDE\footnote{https://archive.codeplex.com/?p=aide} is under the process of active development and has been developed in $C\# .Net$. It is available freely as an open source tool for learning different kinds of automata including DFA, deterministic Mealy machines, NFA and Interface automata (IA). Like other machine learning approaches, automata learning falls into the category of either online/active  or offline/passive approach, and AIDE supports both of these approaches. The online learning setting is based on MAT theory in which the learner communicates with the teacher, called oracle. For a real system where the oracle does not have access to the model, it has to approximate the equivalence queries by testing. For this purpose, AIDE is equipped with a \textit{Testing Engine/testing tool} to make the tool able to perform these kinds of queries. Besides of testing tool, AIDE also provides the facility of \textit{Automata tool} for generating different automata as benchmarks to test the tool and to evaluate different algorithms.



\subsubsection{RALib}
RALib\footnote{https://bitbucket.org/learnlib/ralib} is an open source library and implemented in JAVA for active learning of register automata (a form of extended finite state machines). RALib is an extension of LearnLib \cite{merten2011next} for automata learning and licensed under the \textit{Apache License, Version 2.0}. It provides an extensible implementation to learn models of SULs in the form of register automata in an active way. Besides, it provides modules for output, mixing different tests on data values, typed parameters, and directly inferring models of JAVA classes. RALib also provides heuristics for finding counterexamples as well as a range of performance optimizations (e.g., reducing the length of counterexamples). Sofia Cassel et al. \cite{cassel2015ralib} evaluated RALib on a set of benchmarks and compared it with other tools like \textit{Tomte}\cite{aarts2014tomte,aarts2012automata} and \textit{$LearnLib^{RA}$}(this version has functionality for learning Mealy machines and EFSMs) \cite{howar2012inferringSemantic,howar2012inferringCanndSo} for learning EFSMs and showed that RALib is superior with respect to expressivity, features, and performance. We refer the interested readers to \cite{cassel2015ralib} for further details about the comparison between RALib and other tools that infer register automata. And we refer to \cite{aarts2014algorithms} for viewing the details about the comparison between Tomte and $LearnLib^{RA}$. Results proved the fact that $LearnLib^{RA}$ outperformed Tomte particularly on the smaller models, but on the other hand, Tomte required fewer tests to infer a model completely of larger models (SULs). It provides two tools that can be used from the shell: (i) an IO simulator and (ii) a Java class analyzer. The IO simulator uses a register automaton model as a SUL and can be used to evaluate different algorithms. The class analyzer can be used to infer models of java classes.

\subsubsection{RALT}

RALT\footnote{https://bitbucket.org/muzammil786/ralt} is a reverse-engineering tool, developed in JAVA, to infer the models of real systems taken as a black-box. It conjectures the behavior of SUL in the form finite state machine by generating tests. It outputs finite state models of SUL in two formats: DOT and JFLAP. The resulted model is then viewed and analyzed by a visualizer. Graphviz, for example, is used for visualization of DOT formats. RALT worked under two assumptions. Firstly, the SUL should be a finite state machine and must behave like a Mealy machine. This means that there is an output for every input. Secondly, RALT requires a test driver for running tests on SUL and provides the results of tests back to RALT.




\subsection{Comparison}
Various learning libraries/tools are being used in industry to infer models of realistic systems. A comparison of some important libraries has been presented in the table:\ref{table:Tools Analysis}. 
  





 




\begin{center}

\begin{longtable}{|p{1.2cm}|p{2.3cm}|p{1.3cm}|p{1.8cm}|p{1.0cm}|p{3.0cm}|p{2.0cm}|p{1.0cm}|}

\caption{Overview of the most important learning libraries/tools} 
\label{table:Tools Analysis}\\
\hline 
 
\textbf{TOOLS} & \textbf{Learning Type\newline +Algos}& \textbf{Learning\newline Models}& \textbf{Target\newline Language\newline +Platform} & \textbf{Open Source}&  \textbf{Key Features}&  \textbf{Current\newline Status}& \textbf{Refer-\newline ence}\\ \hline  
\endfirsthead  

\multicolumn{3}{c}%
{{\bfseries \tablename\ \thetable{} -- continued from previous page}} \\ 
\hline 

\textbf{TOOLS} & \textbf{Learning Type\newline +Algos}& \textbf{Learning\newline Models}& \textbf{Target\newline Language\newline +Platform} & \textbf{Open Source}&  \textbf{Key Features}&  \textbf{Current\newline Status}& \textbf{Refer-\newline ence}\\ \hline 
\endhead
 
\hline \multicolumn{8}{|r|}{{Continued on next page}} \\ \hline
\endfoot

\hline
\endlastfoot
 
\hline  
\textbf{Learn-\newline Lib}&Both Active and Passive support; For algorithms see table:\ref{table:Comparison of Different Learning Algorithms}&DFA, NFA, Mealy, VPDA&JAVA; All OS running jre&Yes& Filters, normalizers, visualization, statistics, Logging facilities and Complemented by AutomataLib library&Updated regularly, current version is 0.13.1 (May, 2018)&\cite{learnLibWebSite}\\ 
\hline    
\textbf{LibAlf}&Both Active and Passive support;For algorithms see table:\ref{table:Comparison of Different Learning Algorithms}&DFA, NFA, Mealy, visibly\newline1-counter\newline automaton&C++; MSWindows, Linux and Mac OS&Yes&Filters, statistics, visualization, normalizers and Logging facilities; Also, Complemented by two additional libraries: (i)liblangen (ii)AMoRE++ &No updation, current version is 0.3, last updation was made on April, 2011.&\cite{LibalfWebSite}\\
\hline 
\textbf{AIDE}&Both Active and Passive support;For algorithms see table:\ref{table:Comparison of Different Learning Algorithms}&DFA, NFA, Mealy, IA &$C\#.Net$;MS windows &Yes &Support for: (i)Automata tool (ii) Testing tool &No updation&\cite{AIDEWebSite}\\  
\hline  
\textbf{RALib}&Active Learning; Algo: SL* \cite{cassel2014learning}&EFSMs/RA&JAVA; All OS running jre&Yes&Two tools that can be used from the shell: (i) IO simulator (ii) Java class analyzer; support for multiple theories (testing equalities and inequalities with constants and fresh data values)&Being updated regularly, Current version is 2.0 (August, 2018)&\cite{RALIBWebSite}\\
\hline
\textbf{RALT}&Active Learning; Algos: visit table:\ref{table:Comparison of Different Learning Algorithms}&DFA,Mealy, PFSM, K-Tree&JAVA ; All OS running jre&No&Support for visualization, statistics, logging;&Being updated regularly, Current version is 5.3.3 (December, 2017)&\cite{shahbaz2007learning}\\
\hline
\textbf{TOMTE}&Active Learning&ESFM/RA&JAVA; All OS running jre&Free-ware; Extension of LearnLib&Construction of abstraction fully automatically;Support for visualization, statistics, logging;&Being updated regularly, Current version is Tomte-0.41 (September 2016)&\cite{TomteToolBWebSite}\\
\hline
\textbf{ALEX \footnote{https://learnlib.de/projects/alex/}}&Active Learning&Mealy\newline machine&JAVA ; All OS having Java JRE 8&Free-ware; Extension of LearnLib&Active automata learning for web applications and JSON-based REST services. It has support for visualization, statistics, logging;&Being updated regularly, Current version is 1.5.1 (June, 2018)&\cite{ALEXTool}\\
\hline
\textbf{PSYCO}&Active Learning; it combines dynamic and static analysis techniques&Symbolic behavioral\newline interfaces &JAVA;All OS having Java JRE&Free-ware&Dependencies on jpf-core and jpf-jdart; Support for visualization, statistics&Last updation on Oct, 2016&\cite{PSYCOToolBWebSite} \\
\hline

\end{longtable}
\end{center}

\section{Applications of Model Learning} \label{Sec:Applications of Model Learning}
Currently, model learning is being applied successfully in numerous areas including regression testing, inferring models of existing standardized protocols, compositional reasoning, model-based testing, analysis of inferred models by model checking, etc. In the following subsections, we shall briefly discuss the applications of model learning in different domains. 

\subsection{Network Protocols}
Today we are completely dependent and relying more and more on the reliability of network and security protocols including  SSH, TLS, TCP/IP, EMV, Bluetooth, etc to protect our information and communications. Bugs or vulnerabilities in protocol specification or implementation may cause catastrophic loss. Model learning (or in general, reverse engineering of protocols automatically) expose and/or mitigate such problems effectively. Inferring behavioral models of protocols is important for the understanding of these systems as well as for model checking and model-based testing. \citet{antunes2011reverse} proposed a novel approach for inferring protocol specifications automatically from a log of network traces. Their approach is mainly for clear-text protocols which are used on network servers, for instance, HTTP, IMAP, SIP, FTP, and Microsoft Messenger. In this approach, there is no need to access protocol implementation or source code, and hence it is suitable for open or closed protocols.\par

A process of analysis for understanding the structure and functionality of a device or system is called reverse engineering. In protocols domain, reverse engineering takes the protocol as a black box and derives its application-level specification.  Paolo Milani et al. \cite{comparetti2009prospex} presented a \textit{Prospex}, a system that derives specifications of protocol automatically at the application layer. The proposed system Prospex monitors the working of a server program  that processes network input. Then on the basis of the recorded log file, the tool generates accurate message  format specifications and a generalized protocol state machine. Deriving protocol specifications manually is time-consuming and a tedious job. Weidong Cui et al. \cite{cui2007discoverer} presented \textit{"Discoverer"}, a tool that reverses engineers the formats of protocol message fully automatically by using the network traces of an application. The authors proved their concept by demonstrating the Discoverer tool. They inferred message formats of three network protocols including a text protocol (HTTP) and two binary protocols (RPC and CIFS/SMB) effectively.\par

Model-based verification and validation utilized model learning effectively and produce models of components by observing the external behavior. Basic learning algorithms such as L* generate a modest-size behavioral model of the system, and it implies that the alphabet must be made finite. Model inference techniques which employ these algorithms suppress message parameters, although these parameters have a significant influence on control flow in many communication protocols. The message parameters can be flags, configuration parameters, sequence numbers, agent and session identifiers, etc. Test generation tools like ConformiQ Qtronic \cite{huima2007implementing} which are being used in model-based testing deal with the data influence on control flow. Fides Aarts et al. \cite{aarts2010generating,aarts2015generating} proposed a framework that modifies the learning technique so that it can have data parameters in messages and states for the generation of components with large message alphabets. In this approach, the concept of predicate abstraction was employed to infer the model of SIP protocol. The authors implemented their technique using LearnLib tool and interfaced it with protocol simulator ns-2. They learned the SIP model successfully from the implementation of ns-2.\par 

Learning algorithm like L* \cite{Angluin1987} works only with a small number of abstract inputs and outputs. TCP protocol, like other protocols, use parameters in messages. It also makes use of several remember variables for maintaining a connection. Fides Aarts \cite{aarts2015generating} used the concept of mapper component which was introduced by Aarts, Jonsson, and Uijen. The mapper component abstracts several TCP packets into small abstract actions which are easily handled by the learning tool like LearnLib. They learned successfully different implementations of TCP protocol including Windows 8 and Ubuntu 13.10. The author Ramon Janssen \cite{janssen2013learning} also learned a state diagram of TCP protocol using abstraction technique. Fides Aatrs developed a Tomte \footnote{http://www.italia.cs.ru.nl/tomte/} tool \cite{aarts2014tomte} which construct mapper automatically. The author Max Tijssen \cite{tijssen2014automatic} inferred the models of SSH security protocol implementations using model learning techniques. They inferred the models of freeSSHD, OpenSSH and bitvise servers implementations.

\subsection{Model Checking}
Although the learned models through model learning can be analyzed manually \cite{aarts2010inference,aarts2013formal,de2015protocol} but it is a tedious job. Doron Peled was the first one who introduced the idea of combining model learning and model checking and presented the idea with name \textit{black box checking}. Model checking technology can be used effectively for deep analysis of models since the inferred models may be too complex for manual inspection. In one approach, model checking was used for analysis of models that were made manually using protocol standards. The employed approach captured some bugs, but this activity is time-consuming and error-prone. Moreover, protocol implementations often do not exist in accordance with specifications. Detection of implementation-specific bugs or vulnerabilities become difficult by using this approach of model checking. For instance, Paul Fiterau-Brostean et al. \cite{fiteruau2014learning} showed that TCP implementations in Windows 8 and Ubuntu 13.10 violate the standard (RFC 793) specifications. \citet{de2015protocol} inferred and analyzed three implementations of TLS protocol and found security flaws due to violations of the standard. \citet{chalupar2014automated} configured Lego robot and used model learning technique to reverse engineering the implementation embedded into hand-held smart card readers which are used in internet banking. The analysis of the inferred model showed that the implementation violates the standard. \citet{tijssen2014automatic} also reported the violation of standard in the implementation of SSH protocol \cite{tijssen2014automatic}. In fact, all the violations of standard's specifications reported in \cite{fiteruau2014learning,de2015protocol,chalupar2014automated,tijssen2014automatic} have been discovered using state-of-art model learning technique.\par

To analyze TCP implementations, Fiterau-Brostean et al. \cite{fiteruau2016combining} combined both techniques, i.e., model learning and model checking to obtain formal models of Linux, Windows, and FreeBSD TCP server \& client implementations. They analyzed the learned models with nuSMV model checker to check what are the results when these components interact (e.g., a FreeBSD client and a Linux server) with each other. In another study, to analyze the implementations of SSH protocol, Fiterau-Brostean et al. \cite{fiteruau2017model} used model learning in combination with abstraction techniques for inferring state machines models of Bitvise, OpenSSH, and DropBear SSH server. Later on, they applied model checking technology on learned models for further analysis. They selected many security as well as functional properties from SSH RFC specifications and formalized them in LTL. The authors then checked the formalized properties on the obtained models of SSH protocol using nuSMV model checker and found a number of standard violations. Mathijs Schuts et al. \cite{schuts2016refactoring} presented an approach at Philips and used model learning in combination with equivalence checking technologies for improving a new implementation of the legacy control component. They applied model learning to old implementation and at the same time on new implementation. In their study, they used equivalence model checker for the comparison of learned models.\par



\subsection{Testing and Verification}
The testing and formal verification activity of black box software components is a challenging task. One solution to handle this problem is the combining of testing and learning techniques so that the learned models of software components can be used to explore unknown implementation and ease the testing efforts also. In recent years profound progress has been made in this area and now receiving much attention among the community of testing and verification. Fiterau-Brostean et al. \cite{fiteruau2016combining} inferred the models of TCP implementations and in another research work \cite{fiteruau2017model} they inferred models of SSH implementations. They selected some security and functional properties from the RFC specifications and formalized them in linear temporal logic (LTL). They verified these formal properties on the corresponding learned models using nuSMV model checker and in their findings they discovered various standard violations.\par

Fuzz testing (or fuzzing) is a black box testing technique where the system (e.g., a security protocol) being tested is bombarded with a vast amount of random data (test cases), called fuzz, in an attempt to make it crash. The system is then monitored for any vulnerability exposed by this process. A fuzzer, which is a software tool, can be used to analyze potential causes.  Fuzzing technique is being used successfully to discover implementation and security errors in software, networks, and operating systems. De Ruiter et al. \cite{de2015protocol} used model learning technique to infer state machines from protocol implementations. They called this technique as protocol stat fuzzing because it involves fuzzing different sequences of messages. The inferred state machines are then analyzed manually to look for spurious behavior that can be an indication of logical flaws in the program. Verleg et al. \cite{verleg2016inferring} inferred the state models of SSH implementations using protocol state fuzzing. Fiter et al. \cite{fiteruau2017model} also used the same methodology and inferred models of SSH implementations and inspected the inferred models by nuSMV model checker to look for the spurious behavior of security protocol. Somorovsky et al. \cite{somorovsky2016systematic} showed that vulnerabilities might present in the widely used TLS libraries and these vulnerabilities can be exposed by systematic fuzzing. They presented an open source framework called a TLS-Attacker for evaluating TLS libraries as well as a two-stage fuzz testing technique for the evaluation of TLS server behavior.\par

Alex Groce et al. \cite{groce2002adaptive} used a technique for automatic verification from the area of black box testing and machine learning for handling inconsistencies between a system and its model. These inconsistencies may be due to modeling errors or because of recent modifications made in the system. They implemented their technique as AMC (adaptive model checking) which attempted to perform automatic verification despite the presence of such discrepancies in the model or system. Guoqiang Shu et al.\cite{shu2007testing} proposed a novel learning-based technique for automatic testing of protocol implementation security properties. In their approach, they formally defined a symbolic parameterized extended finite state machine (SP-EFSM) model and confidentiality property for protocol under testing. Their learning-based technique adapts and employs a classic supervised algorithm from the domain of machine learning to analyze the structure of black-box implementations. They developed an algorithm that utilized conformance testing techniques to simulate the teacher. Andreas Hagerer et al. \cite{hagerer2002model} presented an approach called \textit{regular extrapolation}. It gives descriptions of systems, as a model, a posteriori in an automatic fashion. Regular extrapolation builds models on the basis of observations by using techniques from the areas of machine learning and finite automata theory. It is well suited for regression testing, where the availability of the previous version of the system can be used as an approximate reference. The authors used this approach in telecommunication systems.\par


The integration of prefabricated third-party components, which are loosely coupled in a distributed architecture, are being used extensively in designing complex systems like telecom services. Due to this, the system integrator faces many problems while integrating such black-box COTS components. Keqin Li et al. \cite{li2006integrationGuInc} addressed the problem of integration testing of COTS. They proposed a technique which combined machine learning algorithms with test generation techniques. Their approach learned I/O models of COTS (with a slightly modified version of Angluin\textquotesingle s algorithm \cite{Angluin1987}) and based upon these inferred models they defined an integration testing procedure. The inferred state models are then utilized for generating tests. In another work, Keqin Li et al. \cite{li2006integration} used the same methodology but focused on richer modeling formalisms that are more expressive for designing complex systems. They proposed a learning algorithm that was able to learn a black-box component in the form of I/O parameterized model (called PFSM \cite{shahbaz2007learning}). Their sole objective was to facilitate integrator who may derive systematic tests to analyze component interactions. Shahbaz in his detailed research work \cite{shahbaz2014analysis,shahbaz2008reverse,shahbaz2007learningTest,shahbaz2007model} also analyzed the utilization of parameterized finite state machine (PFSM) model in the environment where behavioral models of components and documentation are unavailable. In another challenging study regarding verification of a modular system that is consist of communicating components, Roland Groz et al.\cite{groz2008modular} combined three techniques, i.e.,  inference, testing, and reachability. They addressed the problem of discovering intermittent errors in the situation when models are not available. They used reachability analysis on inferred models to find intermittent errors and compositional problems.\par


Tools that generate tests for formal state machine based specifications \cite{lee1996principles} are also useful for integration testing. Neil Walkinshaw et al. \cite{walkinshaw2010increasing} also addressed the challenging job of test generation to achieve functional coverage, in the environment where complete specification was missing. The inductive testing technique makes use of tests to explore the system behavior and then construct a model in the light of received responses. The inferred model is then further used for test generation. They highlighted the fact that inductive testing is much better than conventional non-inductive strategies by applying them to realistic black-box systems. The authors showed that inductive testing is more efficient than classical black-box techniques for test generation. In another study regarding test generation, Harald Raffelt et al. \cite{raffelt2007dynamic} talked about dynamic testing that utilized a model learning technique for testing of black-box systems. Symbolic execution is a well-known technique and used for test generation of systems having source code access (white-box scenario) \cite{king1976symbolic}. However, this technique faced difficulties in dealing with third-party black-box components that had no access to source code. In such a scenario, behavioral models obtained through model learning techniques enable symbolic execution more accurate and efficient \cite{zhang2015regular}.

\subsection{Compositional Reasoning/Verification}
Model learning is also contributing a lot to the promising domain of compositional reasoning and verification. In compositional reasoning, we treat each software component (like a library function or a concurrent thread) in isolation without any knowledge of software context (like rest part of the software or any other environment thread) where it will be installed. Compositional verification presents an efficient way of handling state explosion problem associated with model checking. In compositional verification, by following the approach of "divide and conquer", the properties of the whole system are break-down into the properties of its components. In this way, if all components satisfy their corresponding properties, then it means that the whole system is satisfied. However, components which are being model checked in separation may satisfy properties in a specific context/environment. This thing generates the requirement for the assume-guarantee style of reasoning \cite{cobleigh2003learning,puasuareanu2008learning}. It is a compositional technique for improving the scalability of model checking\par

A number of theoretical frameworks exist for assume-guarantee reasoning, but they are less practical due to the involvement of non-trivial human interaction. \citet{cobleigh2003learning} presented a framework for performing assume-guarantee reasoning in an incremental and fully automated way. To analyze a component against a property, their approach produced assumptions using L* learning algorithm in combination with model checking (for counterexamples) that was the requirement of environment for satisfying the property to hold. If the property holds then the process will return true with a guarantee of termination, and if not successful then a counterexample will be returned. They implemented their technique in LTSA tool and applied successfully to NASA system. Model learning was employed in another work performed by \citet{he2016learning}, where the authors presented a learning-based assume-guarantee regression verification technique. In model checking, much internal information is computed during the running process of verification. Also, the two consecutive revisions have some common behaviors which may be utilized in the verification process. For example, when one revision is completed, then the internal information computed by model checking may still be useful for the verification of the next revision. \citet{he2016learning} fully utilized this idea in their technique and they reuse the contextual assumptions of the previous round of verification in the current round. Similarly, there are some other case studies where model learning algorithms were utilized for learning assumptions for compositional verification automatically \cite{chen2010comparing,chen2010automated,he2014symbolic}.


\subsection{Miscellaneous Applications}
Apart from the above mentioned applications of model learning there are numerous other cases where it has been used successfully. The range includes, detection of malware \cite{xiao2017automatic}, application and web-services \cite{windmuller2013active,bertolino2009automatic}, in the setting of CTI systems \cite{margaria2002system, niese2001library}, testing brake-by-wire system \cite{feng2013case}, minimizing partially specified systems \cite{oliveira2001efficient}, GUI testing \cite{choi2013guided}, control software of printers \cite{smeenk2012applying}, fighting bot-nets \cite{cho2010inference}, typestate analysis \cite{alur2005synthesis,xiao2013tzuyu}, bank cards \cite{aarts2013formal}, smart-card reader \cite{chalupar2014automated}, Biometric passport \cite{aarts2010inference}, learning Java programs \cite{walkinshaw2007reverse} and many more.


\subsection{Comparison}
\begin{center}
 
\begin{longtable}{|p{2.7cm}|p{4.0cm}|p{2.0cm}|p{2.6cm}|p{2cm}|p{2cm}|}
\caption{Comparison of Applications}
\label{table:Comparison of Applications of Model Learning}\\
\hline

\textbf{Application}  &  \textbf{Purpose} & \textbf{Active/\newline Passive\newline Learning}& \textbf{Learned\newline Model} & \textbf{Library/\newline Tool} & \textbf{Reference} \\ \hline  
\endfirsthead       
    
\multicolumn{3}{c}%
{{\bfseries \tablename\ \thetable{} -- continued from previous page}} \\ 
\hline  
  
\textbf{Application}  &  \textbf{Purpose} & \textbf{Active/\newline Passive\newline Learning}& \textbf{Learned\newline Model} & \textbf{Library/\newline Tool} & \textbf{Re ference} \\ \hline
\endhead 

\hline \multicolumn{6}{|r|}{{Continued on next page}} \\ \hline
\endfoot

\hline
\endlastfoot

\hline 
\textbf{Clear-text Protocols}& To derive protocol specifications of HTTP, SIP, IMAP, FTP, and Microsoft Messenger &Passive&FSM also called \enquote{protocol state machine}& \enquote{ReverX}; Protocol specification inferring tool&\citet{antunes2011reverse}\\ 
\hline 
\textbf{Protocol Specifications Extraction}&Reverse-Engineering of application level protocols (SMB, SMTP, SIP)&Passive&FSM (\enquote{protocol state machine})&\enquote{Prospex}; a system to extract protocol specifications&Paolo Milani et al. \citet{comparetti2009prospex}\\ 
\hline   
\textbf{Formal Verification of SSH Security Protocol}& Analysis of different implementations of SSH.&Active& FSM&LearnLib&\citet{tijssen2014automatic}\\
\hline
\textbf{Learning Protocol Message Format}&Reverse-Engineering of HTTP, PRC, CFIS/SMB protocols&Passive&FSM (\enquote{protocol state machine})&\enquote{Discoverer};\newline a tool& \citet{cui2007discoverer}\\
\hline
\textbf{Inferring model of SIP Protocol}&To check the conformance of SIP implementation with RFC specifications&Active&FSM&LearnLib;\newline ns-2 simulator and Tomte&Fides Aarts et al. \cite{aarts2010generating,aarts2015generating}\\  
\hline
\textbf{Learning TCP Network Protocol}&Analysis of TCP protocols in Windows 8 and Ubuntu 13.10 implementations&Active&FSM&LearnLib&\citet{fiteruau2014learning} + \newline \citet{janssen2013learning}\\ 
\hline
\textbf{Formal Verification of TLS Protocol} (Protocol State Fuzzing)& Analysis of TLS implementations  &Active&FSM&LearnLib&\citet{de2015protocol}\\
\hline 
\textbf{Smartcard Readers}\newline (Reverse engineerings)& To analyze the implementation of smartcard reader (e.dentifier2) using Lego robot.&Active&FSM&LearnLib&\citet{chalupar2014automated}\\
\hline   
\textbf{Security Evaluation of Banking Cards }& To analyze the implementation of EMV protocols embedded in banking cards &Active&FSM&LearnLib&\citet{aarts2013formal}\\
\hline  
\textbf{Analysis of Biometric Passport}& Inferring and analyzing the state machine of biometric passport.&Active&FSM&LearnLib&\citet{aarts2010inference}\\
\hline  
\textbf{Formal Verification of TCP Implementations}& Analysis of different implementations of TCP&Active;\newline FSM&LearnLib&nuSMV model checker&\citet{fiteruau2016combining}\\
\hline  
\textbf{Analysis of SSH implementations}& To analyze different SSH implementations&Active;\newline FSM&LearnLib&nuSMV&\citet{fiteruau2017model}\\ 
\hline 
\textbf{Legacy Software}& Application of model learning in legacy software component&Active&FSM&LearnLib; Equivalence model checker&\citet{schuts2016refactoring}\\
\hline
\textbf{Testing Security Properties} (Learning based testing)& To test security properties of protocol implementations&Active&SP-EFSM&Use of learning algorithm&\citet{shu2007testing}\\
\hline
\textbf{Integration Testing} (learning based testing)& To address the problem of integration testing of COTS.&Active&FSM&Learning algorithm&\citet{li2006integration}\\
\hline
\textbf{Integration Testing} (learning based testing)& To address the problem of integration testing of COTS that are black box components&Active&PFSM&Learning algorithm&\citet{li2006integration}\\
\hline
\textbf{Integration Testing} (learning based testing)& To address the problem of integration testing of COTS (black box components)&Active&PFSM&Learning algorithm &\cite{shahbaz2014analysis,shahbaz2008reverse,shahbaz2007learningTest,shahbaz2007model,groz2008modular}\\
\hline
\textbf{Learning Based Test Generation}& Application of inductive testing to realistic black-box systems&Active&FSM&Learning algorithm&\citet{walkinshaw2010increasing}\\
\hline
\textbf{Learning Based Test Generation}& Using automata learning to systematically test black box systems.(dynamic testing) &Active&FSM&LearnLib&\citet{raffelt2007dynamic}\\ 
\hline
\textbf{Malware Detection}& To apply model learning technique in malware detection &Active&FSM&Learning algorithm&\citet{xiao2017automatic}\\
\hline
\textbf{GUI Testing}& Guided GUI testing of Android Apps with minimal restart and approximate learning &Active&Extended deterministic labeled transition systems (ELTS) &SwiftHand&\citet{choi2013guided}\\
\hline
\textbf{Control Software of Printers/Copiers}&Application of automata learning to industrial control software&Active&FSM&LearnLib&\cite{smeenk2012applying,Smeenk2012ApplyingAL}\\
\hline
\textbf{Botnet Command and Control} (c\&c) protocols&To analyze botnet command and control protocols&Active&FSM (protocol state machine)&Use of learning algorithm&\citet{cho2010inference}\\
\hline    
\textbf{Learning Stateful Typestates}& To models of stateful typestates&Active&FSM&TzuYu tool&\cite{alur2005synthesis,xiao2013tzuyu}\\
\hline 
\textbf{Testing IoT Communication}& Model-Based testing of IoT communication using model learning &Active& FSM&Use of a learning algorithm&\citet{Tappler2017ModelBasedTI}\\
\hline 
\textbf{Program verification}& Software and compositional model checking &Active/Passive&state machine model&Use of a learning algorithm&\cite{jhala2009software,mcmillan1989compositional}\\
\hline 
\textbf{Program testing}& Use of learned model by symbolic execution (White-box testing). &Active&state machine model&Use of a learning algorithm&\citet{zhang2015regular}\\
\hline 
\textbf{Learning interface specifications}& Learning interface specifications for Java classes&Active&state machine model&L* learning algorithm, model checker&\citet{alur2005synthesis}\\
\hline 
\textbf{Compositional Reasoning and Verification}& Using model learning to learn assumptions for compositional verification&Active&DFA&L* learning algorithm&\cite{cobleigh2003learning,puasuareanu2008learning,he2016learning,chen2010comparing,chen2010automated,he2014symbolic}

\end{longtable}
\end{center} 

\section{Achievements, Challenges and Future Work} \label{Sec:Achievements Challenges Future Work}
The overarching goal of model learning is to provide state of the art learning techniques for the construction of accurate and reliable models which are utilized in formal validation techniques. In the following subsections, we shall discuss the recent advancements, challenges and future goals in the domain of model learning.

\subsection{Achievements:} \label{sec:Achievements}
During recent years, tremendous progress has been made in the field of model learning, and table \ref{table:Comparison of Applications of Model Learning} presents its successful applications in multidisciplinary areas of real life. The significant progress made in learning \& testing algorithms, model inferring formalisms, advancements in learning techniques and in learning tools have played a crucial role in scaling the applications of model learning to realistic systems.\par


We first discuss the achievements made in the field of algorithms development. Dana Angluin \citet{Angluin1987} introduced L* algorithm to learn model in the form of DFA, which was later extended by Niese \cite{hagerer2002model} to model Mealy machines. Shahbaz \&Groz \cite{shahbaz2009inferring} proposed improvements in the existing algorithm which resulted in the reduction of worst-time learning complexity. Kearns \& Vazirani \cite{kearns1994introduction} also improved L* algorithm by replacing the observation table by discrimination tree which reduces the number of MQs. In case of counterexample found, the learner, i.e., L* adds all the prefixes of a counterexample in rows of the table and again start posing MQs to refine the hypothesis. Counterexample generator may produce a long, not minimal counterexamples which resulted in posing of numerous redundant MQs. Rivest \& Schapire \cite{rivest1993inference} improved the L* algorithm and instead of row addition, they add an appropriate suffix as a column. \citet{isberner2014ttt} proposed TTT algorithm that is considered to be more efficient. The differentiating characteristic of this algorithm is related to the organization of observations which is redundancy free and can be used to achieve optimal (linear) space complexity. TTT algorithm stores data in data structures of tree-shaped which leads to a compact representation. It is well suited in the context of runtime verification where counterexamples may be very long. We referred the interested readers to \cite{isberner2014ttt} for more details about the comparison of TTT with other learning algorithms.\par

To speed up the learning process, some techniques have been developed including parallelization and checkpointing proposed by \citet{henrix2015performance}. Learning tool like LearnLib is a single threaded, which means that if we would have access to a large computer cluster, then the required execution time (or learning time) will remain about the same. The total learning time can be minimized by running multiple instances of the SUL at the same time. Hence, the total amount of work done remains unchanged, but by distributing it (parallelization), the work can be done in less time (reduced learning time). Learning time has also been improved by checkpointing, i.e., by saving and restoring software states.\par

Basic learning algorithms like L* are successful only for small state machines (DFA, MM). For realistic systems, we have to look for some richer modeling formalisms along with corresponding efficient learning algorithms, and some easy mechanism for construction of abstraction/mapper component. In recent years, efforts have been made for generalization of learning algorithms to richer modeling formalisms like EFSMs. These models have more expressive power as compared to simple state machines and have the ability to store, process and communicate data values. One particular type of richer models is \textit{RA} (see section \ref{sec:EFSM and RA}), and a number of learning algorithms have been developed for this modeling formalism.  In register automaton, actually, states are further extended with register variables that can hold data values. By using these data values as parameters in input/output actions, one may test their equality in transition guards \cite{Vaandrager2017}. Figure:\ref{fig:A_Register_Automaton} presents a simple example of a register automaton for a FIFO-set with capacity two. Interested readers are referred to  \cite{vaandrager2012active,aarts2014algorithms} for further details.\par

\citet{aarts2015learning} proposed an approach for learning RA. They implemented their approach in Tomte software tool that utilized CEGAR technique for the construction of mapper automatically. \citet{cassel2016active} proposed an algorithm for learning realistic systems in the form of EFSM formalism. Their learning algorithm infers register automata models (having variables, guards, and operations) fully automatically which based on a special theory allowing a set of operations and tests on the data domain that can be used in guards. The algorithm has been implemented in LearnLib \cite{isberner2015open} and RALib \cite{cassel2015ralib}. We need robust and versatile implementations of learning algorithms along with surrounding infrastructure for the application of model learning in practical life. In this regard, a great amount of work in the form of learning libraries has been done. Most of these libraries are free, open source and implemented in JAVA. These libraries provide a sophisticated set of algorithms both for active and passive learnings for various automata models along with a variety of equivalence checking algorithms. The main objective of these learning tools is to provide a framework for research on automata learning algorithms and the applications of these algorithms in real life. We referred the interested readers to section \ref{sec:Tools} for more details.


\subsection{Current Challenges} \label{sec:Current Challenges}
No doubt, significant progress has been made in model learning domain, as highlighted in last subsection \ref{sec:Achievements}, but the field is still in developing phase and getting maturity with the passage of time. Model learning techniques required more attention to bring it from current academic level to that where it can be easily and readily applied to real systems. Now we discuss some main challenges which are being faced by model learning technique.\par

First one is the limited expressibility of various modeling formalisms. For example, a Mealy machine is a good option for modeling reactive systems, and in Mealy deterministic formalism, there is one output corresponding to one input (section: \ref{sec:Mealy machine}). In practice, however, the realistic system may have zero or more outputs corresponding to one input. It may also be possible that the behavior of the system is time-dependent, e.g., the output may only occur if the corresponding input is applied for a certain amount of time, etc. The Mealy machine model does not handle such behaviors of realistic systems due to its low power of expressiveness. In one research work, for example, \citet{fiteruau2016combining} eliminated timing-based behavior to squeeze TCP implementations into Mealy machines. It is true that the recent advancements in learning richer classes of models, i.e., RA is a marvelous achievement that makes model learning applicable to a wide range of systems. However, due to the restrictions imposed on data that no operations are allowed, this wide range of systems becomes small and mostly limited to academic examples. Learning algorithms for RA can be extended to EFSM models, having guards that may contain predicates, using SMT solving \citet{cassel2016active}. But, learning algorithms modeling EFSMs with predicates and operations still, demand great effort \cite{Vaandrager2017}.\par

In practice, we often encounter systems which are deterministic but sometimes may also exhibit non-deterministic behavior. \citet{aarts2010inference} applied model learning technique to infer a model of a biometric passport. According to the specification of the passport, its implementation should be deterministic, but the authors observed nondeterministic behavior also during the learning process. Learning of real-world systems with large (or infinite) alphabets, non-determinism, timing constraints, data as well as control flow, etc. are challenging tasks. Currently, most of the well-known model learning tools, as shown in table \ref{table:Tools Analysis} have not integrated other powerful modeling formalisms given in section \ref{sec:miscellaneous formalisms}. \par

Predicting the correctness of inferred models is another challenge that model learning technology still has to be dealt with. As a model learning technique produces a learned model by posing tests that are finite in number, so we can never be sure about the reliability of the learned model. Using an upper bound on the number of states of the SUL one may employ the traditional testing technique, i.e., conformance testing that provides a test suite assuring the correctness of the learned model. However, this testing strategy is also not feasible because the required number of test queries grows exponentially with the number of states of SUL \cite{Vaandrager2017}. So, the challenge, therefore, is to devise some mechanism that can describe quantitatively about the quality of the learned model. The combined power of black-box and white-box learning techniques can be beneficial for many applications however to combine them is also a challenging job. Among various advantages achieved by combining these two techniques, one is the use of white box methods like static analysis and concolic testing to answer the EQs posed by black box learner.

\subsection{Possible Future Directions}


In this section, we shall point out some potential works to be incorporated in model learning which will enhance its capability so that it can be effectively and readily applied to real life systems. \par

The first one is related to the need for improvements in abstraction techniques. As the current versions of different learning algorithms can only model those systems that test the equality of data parameters but operations on data parameters are not allowed. These algorithms model the SUL in the form of RA, which is a big breakthrough in model learning but its scope is still limited due to the restrictions imposed on operations on data. So, one possible research work in model learning is to extend the current abstraction/mapper techniques so that it allows operations on data and also allow other comparisons like "greater than" or "less than" in guard statements.\par


The second future work is related to learning tools. From table \ref{table:Tools Analysis}, we observed that various learning tools especially LearnLib has contributed a lot and still striving to overcome the challenges in the field of model learning. One of the next future potential work is to enhance the capability of learning tools so that they can learn the behaviors involving timing constraints, non-determinism, modeling data, and control flow, handling of large (or infinite) alphabets, etc. of realistic systems in the form of models given in section \ref{sec:miscellaneous formalisms}. \par
 
The third possible work is related to the developments of equivalence checking algorithms. \citet{aarts2014tomte} concluded in his research work that test selection and coverage are still a big barrier in development and application of active learning tools. One possible future work in this direction is to conduct research work to enhance testing techniques for equivalence approximation especially for non-deterministic systems so that we may be able to get a qualitative model. \par

Fourth possible future work is related with the combining of white-box and black-box learning techniques. Combining both the techniques can be propitious for obtaining the models of specific systems. For instance, the combination of active and passive learning approaches \cite{dallmeier2010mining} can be useful, by first constructing a model using passive learning technique and then refining it with the help of active learning technique or during the active learning, phase uses the previous passively inferred model as an oracle \cite{aarts2014tomte}. So combining model inference techniques is another potential future work in the domain of model learning.

\section{Conclusion}
The quality of software systems especially the correct functioning of safety and mission-critical systems is of great concern and demands effective approaches that complement the existing testing and verification techniques by addressing their shortcomings. In this regard, we surveyed an emerging technique called model learning which is highly effective to explore the hidden structures of black-box systems thoroughly. We focused and reviewed model learning framework, techniques, algorithms, tools and its applications in multidisciplinary domains. Table \ref{table:Comparison of Different Learning Algorithms} surveyed some well-known learning algorithms, their supported modeling formalisms and key features for inferring models of realistic systems. To ease the learning process and to apply it on realistic systems, table \ref{table:Tools Analysis} reviewed different model learning libraries.\par

No doubt, the state of the art model learning technique is promising for real-life systems as proved by our survey table \ref{table:Comparison of Applications of Model Learning}, but the field is still in its early phase of development and facing some challenges also. To bring this technique to a level where it can be readily applied to daily life systems, research work regarding modeling formalisms for richer models, algorithms development for learning, counterexamples generation and testing of richer models, utilization of abstraction techniques and feature-rich learning libraries is required. From the analysis of our comparison summaries presented in tables form, we conclude that model learning is efficient, emerging and promising technique and can complement existing testing and verification techniques by providing learned models of black box systems fully automatically.

\section*{Acknowledgement}

We would like to gratitude Mr. Markus Frohme TU Dortmund for valuable discussions and generous support on LearnLib. I am also very thankful to Mr. Naeem Irfan and Mr. Kashif Saghar for fruitful discussions on the topic of model learning. This work is supported by the National Natural Science Foundation of China (NSFC) under Grant No. 61872016 and National Key Research and Development Program of China under Grant No.2016YFB1000804.

\section*{References}

\bibliography{mybibfile}

\end{document}